\documentclass[aps, prb, twocolumn, superscriptaddress]{revtex4-2}
\usepackage{amsmath, amsfonts, amssymb, ascmac, mathtools, bm, braket, comment, tensor}
\usepackage{multirow}
\usepackage{graphicx}
\usepackage{float, color, xcolor}
\usepackage{siunitx}

\usepackage{bbm}
\usepackage{tabularx}

\usepackage[
pagebackref=false,
colorlinks=true,
linkcolor=blue,
urlcolor=blue,
filecolor=black,
citecolor=red,
pdfstartview=FitV,
pdftitle={},
pdfauthor={},
pdfsubject={},
pdfkeywords={},
pdfpagemode=None,
bookmarksopen=true
]{hyperref}

\usepackage{cleveref}
\providecommand{\confluenthypergeometricjack}[1]{\tensor*[_1]{F}{_1^{(#1)}}}
\providecommand{\tricomihypergeometricjack}[1]{\Psi^{(#1)}}
\providecommand{\gausshypergeometricjack}[1]{\tensor*[_2]{F}{_1^{(#1)}}}
\providecommand{\hypergeometricparams}[3]{\left(\begin{array}{cc}#1 \\ #2\end{array}; #3\right)}

\newcommand{\ii}{\text{i}}

\newcommand{\Z}{$\mathbb{Z}$}
\newcommand{\Ztwo}{$\mathbb{Z}_2$}
\newcommand{\twoZ}{$2\mathbb{Z}$}
\newcommand{\ZoplusZ}{$\mathbb{Z} \oplus \mathbb{Z}$}
\newcommand{\ZtwooplusZtwo}{$\mathbb{Z}_2 \oplus \mathbb{Z}_2$}
\newcommand{\twoZoplustwoZ}{$2\mathbb{Z} \oplus 2\mathbb{Z}$}

\newcommand{\Czero}{$\mathrm{U} \left( 2n \right)/\mathrm{U} \left( n \right) \times \mathrm{U} \left( n \right)$}
\newcommand{\Cone}{$\mathrm{U} \left( n \right)$}
\newcommand{\Rzero}{$\mathrm{Sp} \left( 2n \right)/\mathrm{Sp} \left( n \right) \times \mathrm{Sp} \left( n \right)$}
\newcommand{\Rone}{$\mathrm{U} \left( 2n \right)/\mathrm{Sp} \left( n \right)$}
\newcommand{\Rtwo}{$\mathrm{O} \left( 2n \right)/\mathrm{U} \left( n \right)$}
\newcommand{\Rthree}{$\mathrm{O} \left( n \right)$}
\newcommand{\Rfour}{$\mathrm{O} \left( 2n \right)/\mathrm{O} \left( n \right) \times \mathrm{O} \left( n \right)$}
\newcommand{\Rfive}{$\mathrm{U} \left( 2n \right)/\mathrm{O} \left( 2n \right)$}
\newcommand{\Rsix}{$\mathrm{Sp} \left( n \right)/\mathrm{U} \left( n \right)$}
\newcommand{\Rseven}{$\mathrm{Sp} \left( n \right)$}

\providecommand{\tr}{\operatorname{tr}}

\begin{document}

\title{Field theory of non-Hermitian disordered systems}

\author{Ze Chen}
\thanks{The authors are listed in alphabetical order.}
\affiliation{Department of Physics, Princeton University, Princeton, New Jersey 08544, USA}

\author{Kohei Kawabata}
\thanks{The authors are listed in alphabetical order.}
\affiliation{Institute for Solid State Physics, University of Tokyo, Kashiwa, Chiba 277-8581, Japan}

\author{Anish Kulkarni}
\thanks{The authors are listed in alphabetical order.}
\affiliation{Department of Physics, Princeton University, Princeton, New Jersey 08544, USA}

\author{Shinsei Ryu}
\thanks{The authors are listed in alphabetical order.}
\affiliation{Department of Physics, Princeton University, Princeton, New Jersey 08544, USA}

\date{\today}

\begin{abstract}
The interplay between non-Hermiticity and disorder gives rise to unique universality classes of Anderson transitions.
Here, we develop a field-theoretical description of non-Hermitian disordered systems based on fermionic replica nonlinear sigma models.
We classify the target manifolds of the nonlinear sigma models across all the 38-fold symmetry classes of non-Hermitian systems and corroborate the correspondence of the universality classes of Anderson transitions between non-Hermitian systems and Hermitized systems with additional chiral symmetry.
We apply the nonlinear sigma model framework to study the spectral properties of non-Hermitian random matrices with particle-hole symmetry.
Furthermore, we demonstrate that the Anderson transition unique to nonreciprocal disordered systems in one dimension, including the Hatano-Nelson model, originates from the competition between the kinetic and topological terms in a one-dimensional nonlinear sigma model.
We also discuss the critical phenomena of non-Hermitian disordered systems with symmetry and topology in higher dimensions.
\end{abstract}

\maketitle

\tableofcontents

%%%%% Introduction %%%%%
\section{Introduction}

Disorder plays a fundamental role in condensed matter physics.
One of the remarkable phenomena induced by disorder is Anderson localization~\cite{Anderson-58}, which is disorder-induced localization of coherent waves and significantly influences transport phenomena of solids~\cite{Lee-review, Beenakker-review-97, *Beenakker-review-15, Evers-review}, as well as light~\cite{Segev-review} and cold atoms~\cite{Billy-08, Roati-08}.
The competition between disorder and other system parameters also causes continuous transitions between 
localized and delocalized phases (i.e., Anderson transitions).
A general understanding of Anderson localization and transitions was provided by the scaling theory based on the one-parameter-scaling hypothesis~\cite{Thouless-review, Abrahams-79, *Anderson-80}.
This phenomenological scaling theory was formalized by the perturbative analysis of the renormalization-group beta function~\cite{Gorkov-79, Hikami-80, Altshuler-80}.
Notably, it was also corroborated by an effective field theory description.
Specifically, nonlinear sigma models 
based on the replica~\cite{Wegner-79, *Schafer-80, Efetov-80, Altland-Simons-textbook}, supersymmetry~\cite{Efetov-83, *Efetov-textbook}, and Keldysh~\cite{Kamenev-textbook} approaches
effectively capture the universal behavior of Anderson transitions. 
It is also notable that the zero-dimensional version of the nonlinear sigma models characterizes random matrix theory and underlies the foundation of quantum chaos~\cite{Haake-textbook}.

The universality classes of Anderson transitions are determined solely by spatial dimensions, symmetry, and topology.
Despite the absence of delocalization in one and two dimensions without symmetry~\cite{Abrahams-79}, time-reversal symmetry in the presence of spin-orbit interactions enables delocalization even in two dimensions~\cite{Hikami-80}.
Additionally, chiral (sublattice) symmetry enables delocalization of zero modes even in one dimension~\cite{Dyson-53}.
These fundamental internal symmetries---time-reversal symmetry~\cite{Wigner-51, *Wigner-58, Dyson-62}, chiral symmetry~\cite{Gade-91, *Gade-93, Verbaarschot-94, *Verbaarschot-00-review}, and particle-hole symmetry~\cite{AZ-97}---constitute the tenfold Altland-Zirnbauer symmetry classes~\cite{AZ-97}.
Correspondingly, the target manifolds of the nonlinear sigma models are universally classified in this tenfold way, as summarized in Table~\ref{tab: Hermitian}.
The beta functions for the tenfold symmetry classes were also calculated perturbatively~\cite{Hikami-81, Wegner-89}.

Topology is another crucial ingredient that determines the universality classes of Anderson transitions.
Prime examples include the quantum Hall transition~\cite{QHE-textbook, Huckestein-review, Kramer-review}, in which the nontrivial topology of wave functions induces Anderson transitions even in two dimensions.
Such topological Anderson transitions also accompany topological terms in the effective nonlinear sigma models, leading to the two-parameter scaling based on the longitudinal and Hall conductivity~\cite{Khmel'nitskii-83, Levine-83, *Pruisken-84, *Pruisken-88}.
Beyond the quantum Hall transition, the role of topology in Anderson transitions has been widely studied~\cite{Ludwig-94, Fendley-01, Ostrovsky-07, Ryu-07, *Nomura-07, Bardarson-07, Ryu-12, Fu-12, Kobayashi-13, *Kobayashi-14, MondragonShem-14, Altland-14, *Altland-15, Liu-16, Sbierski-21, Xiao-23, *Zhao-24}.
Depending on spatial dimensions and symmetry, a variety of topological terms have been shown to appear. 
Classification of such possible topological terms also underlies the tenfold periodic table of topological insulators and superconductors~\cite{Schnyder-08, *Ryu-10, Kitaev-09, CTSR-review}.

Recently, beyond the Hermitian regime, the physics of Anderson localization in non-Hermitian systems has attracted growing interest~\cite{Hatano-Nelson-96, *Hatano-Nelson-97, Efetov-97, *Efetov-97B, Feinberg-97, *Feinberg-99, Brouwer-97, Longhi-19, Zeng-20, Tzortzakakis-20, Huang-20, KR-21, Claes-21, Luo-21L, *Luo-21B, Luo-22R, Liu-Fulga-21, Moustaj-22, Nakai-24, Ghosh-23, Liu-24, Huang-24}.
In general, non-Hermiticity arises from the coupling with the external environment and leads to a wide variety of phenomena unique to open systems~\cite{Konotop-review, Christodoulides-review}.
Non-Hermiticity enriches symmetry and changes the
internal symmetry classification from 10-fold to 38-fold~\cite{Bernard-LeClair-02, KSUS-19}.
This is relevant to non-Hermitian random matrix theory and also accompanies characterization of chaotic behavior in open quantum systems~\cite{Grobe-88, *Grobe-89, Xu-19, Hamazaki-19, Denisov-19, Can-19PRL, *Can-19JPhysA, Hamazaki-20, Akemann-19, Sa-20, Wang-20, Xu-21, GarciaGarcia-22PRL, JiachenLi-21, Cornelius-22, GarciaGarcia-22PRX, Prasad-22, Sa-22-SYK, Kulkarni-22-SYK, GarciaGarcia-22PRD, GJ-23, *GJ-24-ETH, Xiao-22, Shivam-22, Ghosh-22, Sa-23, Kawabata-23, Kawabata-23SVD, Xiao-24}.
Topological phenomena intrinsic to non-Hermitian systems have also been studied considerably~\cite{BBK-review, Okuma-Sato-review}.
Notably, non-Hermiticity causes the unique universality classes of Anderson transitions that have no analogs in Hermitian systems. 
In particular, a non-Hermitian extension of the Anderson model with the nonreciprocal hopping, which was first investigated by Hatano and Nelson~\cite{Hatano-Nelson-96, *Hatano-Nelson-97}, exhibits delocalization even in one dimension.
Such an Anderson transition in nonreciprocal disordered systems has no counterparts in Hermitian systems, in which delocalization is forbidden in one dimension without symmetry~\cite{Abrahams-79, *Anderson-80}.
Furthermore, the unique universality classes of Anderson transitions in non-Hermitian systems were shown to be generally related to those of the corresponding Hermitian systems with additional chiral symmetry~\cite{Luo-22R}.

Despite the recent surge of interest in Anderson transitions in non-Hermitian systems, most previous works have hitherto relied on numerical analysis.
This contrasts with the Hermitian case, for which the interplay between the numerical analysis of microscopic models and the effective field theory description leads to a profound understanding.
In Ref.~\cite{Nishigaki-02}, the spectral statistics of non-Hermitian random matrices were investigated on the basis of replica nonlinear sigma models.
Still, the role of symmetry, as well as possible Anderson transitions in finite dimensions, has remained elusive. 
Notably, a scaling theory of nonreciprocal disordered systems in one dimension has recently been developed based on the combination of the conductance from the left to the right and that from the right to the left~\cite{KR-21}, which is reminiscent of the two-parameter scaling of the quantum Hall transition~\cite{Khmel'nitskii-83, Levine-83, *Pruisken-84, *Pruisken-88}.
However, it has been unclear whether this nonunitary two-parameter scaling theory is formulated by 
an effective field theory.

%%%%% Hermitian %%%%%
\begin{table}[t]
	\centering
	\caption{Fermionic replica nonlinear sigma model (NLSM) target manifolds for Hermitian systems.
    The tenfold Altland-Zirnbauer symmetry classes consist of time-reversal symmetry (TRS), particle-hole symmetry (PHS), and chiral symmetry (CS).}
	\label{tab: Hermitian}
     \begin{tabular}{ccccc} \hline \hline
    ~~Class~~ & ~~TRS~~ & ~~PHS~~ & ~~CS~~ & ~~NLSM target manifold~~ \\ \hline
    A & $0$ & $0$ & $0$ & \Czero \\
    AIII & $0$ & $0$ & $1$ & \Cone \\ \hline
    AI & $+1$ & $0$ & $0$ & \Rzero \\
    BDI & $+1$ & $+1$ & $1$ & \Rone \\
    D & $0$ & $+1$ & $0$ & \Rtwo \\
    DIII & $-$ & $+$ & $1$ & \Rthree \\
    AII & $-1$ & $0$ & $0$ & \Rfour\\
    CII & $-1$ & $-1$ & $1$ & \Rfive \\
    C & $0$ & $-1$ & $0$ & \Rsix \\
    CI & $+1$ & $-1$ & $1$ & \Rseven \\ \hline \hline
  \end{tabular}
\end{table}

In this work, we develop an effective field theory description of non-Hermitian disordered systems.
We begin with formulating nonlinear sigma models based on the fermionic replica approach (Sec.~\ref{sec: NLSM}).
Through the Hermitization technique, we classify the relevant target manifolds of the nonlinear sigma models across all the 38-fold symmetry classes of non-Hermitian systems, as summarized in Tables~\ref{tab: complex AZ}-\ref{tab: real AZ + SLS}.
Our classification tables elucidate the universality classes of the spectral statistics and Anderson transitions in non-Hermitian disordered systems.
Specifically, we provide a field-theoretical understanding of the correspondence of the universality classes between non-Hermitian systems and Hermitized systems with additional chiral symmetry, which was formulated for microscopic lattice models in Ref.~\cite{Luo-22R}.

In addition to the general discussion, we illustrate the effect of symmetry by explicitly deriving the nonlinear sigma model from non-Hermitian random matrices with particle-hole symmetry (Sec.~\ref{sec: RMT}).
We also show the duality of the partition function between different symmetry classes, which is a unique feature of non-Hermitian systems.
As an exemplary universality class inherent in non-Hermitian disordered systems, we demonstrate that the Anderson transition in nonreciprocal disordered systems in one dimension, including the Hatano-Nelson model~\cite{Hatano-Nelson-96, *Hatano-Nelson-97}, arises from the competition between the kinetic and topological terms in the one-dimensional nonlinear sigma model (Sec.~\ref{sec: Hatano-Nelson}).
We provide the renormalization-group flow in Fig.~\ref{fig: 1D} and clarify %its physical meaning.
that the topological term manifests itself as a current in contrast to the Hermitian counterparts.
Furthermore, we discuss the possible Anderson transitions in non-Hermitian disordered systems in two and three dimensions, focusing mainly on the consequences of the topological terms (Sec.~\ref{sec: others}).

%%%%%%%%%%
\onecolumngrid

%\clearpage

%%%%% complex AZ %%%%%
\begin{table}[H]
	\centering
	\caption{Fermionic replica nonlinear sigma model (NLSM) target manifolds for non-Hermitian systems in the complex Altland-Zirnbauer (AZ) symmetry classes.
    The complex AZ symmetry classes consist of chiral symmetry (CS) and sublattice symmetry (SLS).
    The subscript of SLS $\mathcal{S}_{\pm}$ specifies the commutation ($+$) or anticommutation ($-$) relation to CS: $\Gamma \mathcal{S}_{\pm} = \pm \mathcal{S}_{\pm} \Gamma$.
    The Hermitized symmetry classes and the possible topological terms in one, two, and three dimensions are also shown.}
	\label{tab: complex AZ}
     \begin{tabular}{cccccccc} \hline \hline
    ~~Class~~ & ~~CS~~ & ~~SLS~~ & ~~Hermitization~~ & ~~NLSM target manifold~~ & ~~$d=1$~~ & ~~$d=2$~~ & ~~$d=3$~~ \\ \hline
    A & $0$ & $0$ & AIII & \Cone & \Z & $0$ & \Z \\
    AIII = A + $\eta$ & $1$ & $0$ & A & \Czero & $0$ & \Z & $0$ \\ \hline
    AIII $+ \mathcal{S}_{+}$ & $1$ & $1$ & AIII & \Cone & \Z & $0$ & \Z \\ \hline
    ~~A $+ \mathcal{S}$ = AIII$^{\dag}$~~ & $0$ & $1$ & ~~AIII $\times$ AIII~~ & \Cone\,$\times$\,\Cone & \ZoplusZ & $0$ & \ZoplusZ \\
    AIII $+ \mathcal{S}_{-}$ & $1$ & $1$ & A $\times$ A & [\Czero]\,$\times$\,[\Czero] & $0$ & \ZoplusZ & $0$ \\ \hline \hline
  \end{tabular}
\end{table}

%%%%% real AZ %%%%%
\begin{table}[H]
	\centering
	\caption{Fermionic replica nonlinear sigma model (NLSM) target manifolds for non-Hermitian systems in the real Altland-Zirnbauer (AZ) symmetry classes.
    The real AZ symmetry classes consist of time-reversal symmetry (TRS), particle-hole symmetry (PHS), and chiral symmetry (CS).
    The Hermitized symmetry classes and the possible topological terms in one, two, and three dimensions are also shown.}
	\label{tab: real AZ}
     \begin{tabular}{ccccccccc} \hline \hline
    ~~Class~~ & ~~TRS~~ & ~~PHS~~ & ~~CS~~ & ~~Hermitization~~ & ~~NLSM target manifold~~ & ~~$d=1$~~ & ~~$d=2$~~ & ~~$d=3$~~ \\ \hline
    ~~AI = D$^{\dag}$~~ & +1 & 0 & 0 & BDI & \Rone & \Z & $0$ & $0$ \\
    BDI & +1 & +1 & 1 & D & \Rtwo & \Ztwo & \Z & $0$ \\
    D & 0 & +1 & 0 & DIII & \Rthree & \Ztwo & \Ztwo & \Z \\
    DIII & -1 & +1 & 1 & AII & \Rfour & $0$ & \Ztwo & \Ztwo \\
    AII = C$^{\dag}$ & -1 & 0 & 0 & CII & \Rfive & \twoZ & $0$ & \Ztwo \\
    CII & -1 & -1 & 1 & C & \Rsix & $0$ & \twoZ & $0$ \\
    C & 0 & -1 & 0 & CI & \Rseven & $0$ & $0$ & \twoZ \\
    CI & +1 & -1 & 1 & AI & \Rzero & $0$ & $0$ & $0$ \\ \hline \hline
  \end{tabular}
\end{table}

%%%%% real AZ-dag %%%%%
\begin{table}[H]
	\centering
	\caption{Fermionic replica nonlinear sigma model (NLSM) target manifolds for non-Hermitian systems in the real Altland-Zirnbauer$^{\dag}$ (AZ$^{\dag}$) symmetry classes.
    The real AZ$^{\dag}$ symmetry classes consist of time-reversal symmetry$^{\dag}$ (TRS$^{\dag}$), particle-hole symmetry$^{\dag}$ (PHS$^{\dag}$), and chiral symmetry (CS).
    The Hermitized symmetry classes and the possible topological terms in one, two, and three dimensions are also shown.}
	\label{tab: real AZ-dag}
     \begin{tabular}{ccccccccc} \hline \hline
    ~~Class~~ & ~~TRS$^{\dag}$~~ & ~~PHS$^{\dag}$~~ & ~~CS~~ & ~~Hermitization~~ & ~~NLSM target manifold~~ & ~~$d=1$~~ & ~~$d=2$~~ & ~~$d=3$~~ \\ \hline
    AI$^{\dag}$ & +1 & 0 & 0 & CI & \Rseven & $0$ & $0$ & \twoZ \\
    BDI$^{\dag}$ & +1 & +1 & 1 & AI & \Rzero & $0$ & $0$ & $0$ \\
    D$^{\dag}$ = AI & 0 & +1 & 0 & BDI & \Rone & \Z & $0$ & $0$ \\
    DIII$^{\dag}$ & -1 & +1 & 1 & D & \Rtwo & \Ztwo & \Z & $0$ \\
    AII$^{\dag}$ & -1 & 0 & 0 & DIII & \Rthree & \Ztwo & \Ztwo & $0$ \\
    CII$^{\dag}$ & -1 & -1 & 1 & AII & \Rfour & $0$ & \Ztwo & \Ztwo \\
    C$^{\dag}$ = AII & 0 & -1 & 0 & CII & \Rfive & \twoZ & $0$ & \Ztwo \\
    CI$^{\dag}$ & +1 & -1 & 1 & C & \Rsix & $0$ & \twoZ & $0$ \\ \hline \hline
  \end{tabular}
\end{table}

%%%%% real AZ + SLS %%%%%
\begin{table}[H]
	\centering
	\caption{Fermionic replica nonlinear sigma model (NLSM) target manifolds for non-Hermitian systems in the real Altland-Zirnbauer (AZ) symmetry classes with sublattice symmetry (SLS).
    The subscript of SLS $\mathcal{S}_{\pm}$ specifies the commutation ($+$) or anticommutation ($-$) relation to time-reversal symmetry (TRS) and/or particle-hole symmetry (PHS).
    For the symmetry classes that involve both TRS and PHS (i.e., classes BDI, DIII, CII, and CI), the first subscript specifies the relation to TRS and the second one to PHS.
    The Hermitized symmetry classes and the possible topological terms in one, two, and three dimensions are also shown.
    Some classes are repeated to clarify the periodicity of the table.}
	\label{tab: real AZ + SLS}
     \begin{tabular}{cccccc} \hline \hline
    ~~Class~~ & ~~Hermitization~~ & ~~NLSM target manifold~~ & ~~$d=1$~~ & ~~$d=2$~~ & ~~$d=3$~~ \\ \hline
    BDI + $\mathcal{S}_{++}$ & BDI & \Rone & \Z & $0$ & $0$  \\
    DIII + $\mathcal{S}_{--}$ = BDI + $\mathcal{S}_{--}$ & DIII & \Rthree & \Ztwo & \Ztwo & \Z \\
    CII + $\mathcal{S}_{++}$ & CII & \Rfive & \twoZ & $0$ & \Ztwo \\
    CI + $\mathcal{S}_{--}$ = CII + $\mathcal{S}_{--}$ & CI & \Rseven & $0$ & $0$ & \twoZ \\ \hline
    AI + $\mathcal{S}_{-}$ = AII + $\mathcal{S}_{-}$ & AIII & \Cone & \Z & $0$ & \Z \\
    BDI + $\mathcal{S}_{-+}$ = DIII + $\mathcal{S}_{-+}$ & A & \Czero & $0$ & \Z & $0$ \\
    D + $\mathcal{S}_{+}$ & AIII & \Cone & \Z & $0$ & \Z \\
    DIII + $\mathcal{S}_{-+}$ = BDI + $\mathcal{S}_{-+}$ & A & \Czero & $0$ & \Z & $0$ \\
    AII + $\mathcal{S}_{-}$ = AI + $\mathcal{S}_{-}$ & AIII & \Cone & \Z & $0$ & \Z \\
    CII + $\mathcal{S}_{-+}$ = CI + $\mathcal{S}_{-+}$ & A & \Czero & $0$ & \Z & $0$ \\
    C + $\mathcal{S}_{+}$ & AIII & \Cone & \Z & $0$ & \Z \\
    CI + $\mathcal{S}_{-+}$ = CII + $\mathcal{S}_{-+}$ & A & \Czero & $0$ & \Z & $0$ \\ \hline
    BDI + $\mathcal{S}_{--}$ = DIII + $\mathcal{S}_{--}$ & DIII & \Rthree & \Ztwo & \Ztwo & \Z \\
    DIII + $\mathcal{S}_{++}$ & CII & \Rfive & \twoZ & $0$ & \Ztwo \\
    CII + $\mathcal{S}_{--}$ = CI + $\mathcal{S}_{--}$ & CI & \Rseven & $0$ & $0$ & \twoZ \\
    CI + $\mathcal{S}_{++}$ & BDI & \Rone & \Z & $0$ & $0$ \\ \hline
    AI + $\mathcal{S}_{+}$ & ~~BDI $\times$ BDI~~ & [\Rone]\,$\times$\,[\Rone] & \ZoplusZ & $0$ & $0$ \\
    BDI + $\mathcal{S}_{+-}$ & ~~D $\times$ D~~& [\Rtwo]\,$\times$\,[\Rtwo] & \ZtwooplusZtwo & \ZoplusZ & $0$ \\
    D + $\mathcal{S}_{-}$ & ~~DIII $\times$ DIII~~ & \Rthree\,$\times$\,\Rthree & \ZtwooplusZtwo & \ZtwooplusZtwo & \ZoplusZ \\
    DIII + $\mathcal{S}_{+-}$ & ~~AII $\times$ AII~~ & [\Rfour]\,$\times$\,[\Rfour] & $0$ & \ZtwooplusZtwo & \ZtwooplusZtwo \\
    AII + $\mathcal{S}_{+}$ & ~~CII $\times$ CII~~ & [\Rfive]\,$\times$\,[\Rfive] & \twoZoplustwoZ & $0$ & \ZtwooplusZtwo \\
    CII + $\mathcal{S}_{+-}$ & ~~C $\times$ C~~ & [\Rsix]\,$\times$\,[\Rsix] & $0$ & \twoZoplustwoZ & $0$ \\
    C + $\mathcal{S}_{-}$ & ~~CI $\times$ CI~~ & \Rseven\,$\times$\,\Rseven & $0$ & $0$ & \twoZoplustwoZ \\
    CI + $\mathcal{S}_{+-}$ & ~~AI $\times$ AI~~ & [\Rzero]\,$\times$\,[\Rzero] & $0$ & $0$ & $0$\\ \hline \hline
  \end{tabular}
\end{table}

\twocolumngrid
%%%%%%%%%%

%%%%% NLSM %%%%%
\section{Nonlinear sigma model}
    \label{sec: NLSM}

We develop a field-theoretical description of non-Hermitian disordered systems:
nonlinear sigma model.
In Sec.~\ref{subsec: Green}, we introduce a formalism to capture statistical properties of non-Hermitian disordered systems, such as the Green's function and characteristic polynomial.
In Sec.~\ref{subsec: replica}, we develop their nonlinear sigma model description based on the fermionic replica theory.
In Sec.~\ref{subsec: classification}, we classify the universality classes of these nonlinear sigma models for all the 38 symmetry classes of non-Hermitian systems, as summarized in Tables~\ref{tab: complex AZ}-\ref{tab: real AZ + SLS}.
In Sec.~\ref{subsec: Hermitian}, we clarify the distinction of the nonlinear sigma models between Hermitian and non-Hermitian systems.

%%%%%%%%%%
\subsection{Green's function}
    \label{subsec: Green}

We consider generic non-Hermitian free fermionic systems,
\begin{equation}
    \hat{H} = \hat{\psi}^{\dag} H \hat{\psi},
\end{equation}
where $\hat{\psi} = (\hat{\psi}_1, \hat{\psi}_2, \cdots, \hat{\psi}_N)$ [$\hat{\psi}^{\dag} = (\hat{\psi}_1^{\dag}, \hat{\psi}_2^{\dag}, \cdots, \hat{\psi}_N^{\dag}$)] denotes $N$ flavors of fermionic annihilation (creation) operators, and $H$ is an $N \times N$ single-particle non-Hermitian Hamiltonian.
While we study zero-dimensional systems (i.e., random matrices) in this section for simplicity, it is straightforward to incorporate the spatial degrees of freedom (see Sec.~\ref{sec: Hatano-Nelson} for a prototypical example of one-dimensional systems).
We assume that $H$ consists of the nonrandom kinetic term $H_0$ and the Gaussian disordered potential $V$,
\begin{equation}
    H = H_0 + V,
\end{equation}
where $V$ satisfies
\begin{equation}
    \braket{V^{ij}} = 0, \quad \braket{V^{ij} (V^{kl})^{*}} = \frac{g^2}{N} \delta^{ik} \delta^{jl}
\end{equation}
with the disorder strength $g \geq 0$.
Later, we take the large-$N$ limit, which enables the saddle-point analysis.
Notably, $H = H_0 + V$ can respect certain symmetry and belong to one of the 38 symmetry classes of non-Hermitian systems (see Appendix~\ref{asec: symmetry} for details)~\cite{Bernard-LeClair-02, KSUS-19}, which further leads to different target manifolds of the nonlinear sigma models.
Below, we first focus on non-Hermitian systems without symmetry (i.e., class A) to illustrate the replica formalism, and shortly discuss the effect of symmetry in a general manner.
Additionally, in Sec.~\ref{sec: RMT}, we explicitly derive a nonlinear sigma model from non-Hermitian random matrices with particle-hole symmetry (i.e., class D) and demonstrate the case of a different target manifold.

To capture the statistical properties of the non-Hermitian disordered systems $H$, we study the Green's function
\begin{equation}
    G \left( E \right) \coloneqq \frac{1}{N} \braket{\mathrm{tr} \left( E-H \right)^{-1}},
        \label{eq: Green}
\end{equation}
where $E$ denotes a complex parameter corresponding to single-particle energy, and the bracket denotes the disorder average.
In principle, $G \left( E \right)$ contains all the physically relevant information of $H$.
As a prime example, from the identity
\begin{equation}
\frac{\partial}{\partial E^{*}} \frac{1}{E} = \pi \delta^2 \left( E \right) \coloneqq \pi \delta \left( \mathrm{Re}\,E \right) \delta \left( \mathrm{Im}\,E \right), 
    \label{eq: delta - 2D}
\end{equation}
the density of states is given as
\begin{equation}
    \rho \left( E \right) \coloneqq \frac{1}{N} \braket{\mathrm{tr}\,\delta \left( E-H \right)} = \frac{1}{\pi} \frac{\partial}{\partial E^{*}} G \left( E \right).
        \label{eq: rho - 2D}
\end{equation}
As a useful quantity, we introduce
\begin{equation}
    \Phi \left( E, E^{*} \right) \coloneqq \braket{\log \det \left( E-H \right) \left( E^{*} - H^{\dag} \right)},
        \label{eq: Phi}
\end{equation}
from which the Green's function $G \left( E \right)$ is obtained as
\begin{equation}
    G \left( E \right) = \frac{1}{N} \frac{\partial}{\partial E} \Phi \left( E, E^{*} \right).
\end{equation}
Here, $\Phi \left( E, E^{*} \right)$ plays a role of the partition function (or free energy) in non-Hermitian disordered systems.
Using $\Phi \left( E, E^{*} \right)$, we also have
\begin{equation}
    \rho \left( E \right) = \frac{1}{\pi N} \frac{\partial^2}{\partial E \partial E^{*}} \Phi \left( E, E^{*} \right) = \frac{1}{4\pi N} \Delta \Phi \left( E, E^{*} \right),
        \label{eq: DOS - potential}
\end{equation}
where $\Delta \coloneqq \partial^2/\partial \left( \mathrm{Re}\,E \right)^2 + \partial^2/\partial \left( \mathrm{Im}\,E \right)^2$ is the Laplacian in the complex plane.
Consequently, the statistical properties of non-Hermitian disordered systems have an analogy with two-dimensional classical electrostatics based on the Coulomb potential $\Phi \left( E, E^{*} \right)$ (see, for example, Ref.~\cite{Sommers-88}).

In addition to the generating function $\Phi(E,E^*)$,
we also consider~\cite{Nishigaki-02}
\begin{equation}
    Z_n \left( E, E^{*} \right) \coloneqq \Braket{\left[ \det \left( E-H \right) \left( E^{*} - H^{\dag} \right) \right]^n}
        \label{eq: replica partition function}
\end{equation}
for $n \in \mathbb{Z}$. 
This can be regarded as
a non-Hermitian analog of 
the characteristic polynomials
in the regular Hermitian random matrix theory.
Assuming the analytic continuation from $n \in \mathbb{Z}$ to $n \in \mathbb{R}$, we can obtain the partition function $\Phi \left( E,E^* \right)$ from the characteristic polynomials $Z_n(E,E^*)$ as 
\begin{equation}
    \Phi \left( E, E^{*} \right) = \lim_{n\to 0} \frac{Z_n \left( E, E^{*} \right)}{n}.
        \label{eq: Phi - ZnE}
\end{equation}
Hence, from $Z_n$, we can derive $\Phi$ and $G$, thereby obtaining physical observables of non-Hermitian disordered systems.
Note, however, that the characteristic polynomials are of interest on their own (see, for example, Refs.~\cite{keating_random_2000, Br_zin_2000}).

%%%%%%%%%%
\subsection{Replica field theory}
    \label{subsec: replica}

In the following, we develop an effective field theory approach to calculate the characteristic polynomials $Z_n(E,E^*)$ in terms of nonlinear sigma models.
The target manifolds of nonlinear sigma models depend on the sign of $n$;
$n > 0$ and $n < 0$ are called the fermionic and bosonic replicas, respectively.
While we focus on the fermionic replicas with $n>0$ in this work, it is significant to formulate the bosonic version of replica nonlinear sigma models in future work.
In passing, we note that a bosonic replica nonlinear sigma model was developed for non-Hermitian random matrices in class AI (see also Appendix~\ref{asec: NH RMT})~\cite{Nishigaki-02}.

To formulate an effective field theory, we use 
\begin{equation}
    \int \left[ \prod_{i=1}^N d\bar{\psi}^i d\psi^i \right] \exp \left[ - \sum_{i, j = 1}^{N} \bar{\psi}^i M^{ij} \psi^j \right] = \det M,
\end{equation}
where $\bar{\psi}^1, \bar{\psi}^2, \cdots, \bar{\psi}^N$ and $\psi^1, \psi^2, \cdots, \psi^N$ are sets of Grassmann variables, and $M$ is a generic $N \times N$ matrix.
Then, $Z_n \left( E, E^{*} \right)$ can be expressed as
\begin{align}
    &Z_n = \bigg\langle \int \mathcal{D} \left[ \bar{\psi}, \psi, \bar{\chi}, \chi \right] \prod_{i, j = 1}^N \prod_{a=1}^{n}\nonumber \\
    &\exp \left[ - \bar{\psi}_a^i \left( E\delta^{ij} -H^{ij} \right) \psi_a^j - \bar{\chi}_a^i \left( E^{*} \delta^{ij} - (H^{\dag})^{ij} \right) \chi_a^j \right] \bigg\rangle,
\end{align}
with the integral measure 
\begin{equation}
\mathcal{D} \left[ \bar{\psi}, \psi, \bar{\chi}, \chi \right] \coloneqq \prod_{i=1}^N \prod_{a=1}^{n} d\bar{\psi}^i_a d\psi^i_a d\bar{\chi}^i_a d\chi^i_a.
\end{equation}
The superscripts describe the original degrees of freedom (i.e., $i= 1, 2, \cdots, N$) while the subscripts describe the replica index $a = 1, 2, \cdots, n$.
Hereafter, we employ the Einstein convention and abbreviate the summation for simplicity.
Owing to the Gaussian distribution, the disorder average can be performed straightforwardly.
In fact, we have
\begin{align}
    Z_n &= \int \mathcal{D} \left[ \bar{\psi}, \psi, \bar{\chi}, \chi, V \right] \exp \left[ - \bar{\psi}_a^i \left( E\delta^{ij} -H^{ij}_0 - V^{ij} \right) \psi_a^j \right. \nonumber \\
    &\quad \left. - \bar{\chi}_a^i \left( E^{*} \delta^{ij} - (H^{\dag}_0)^{ij} - (V^\dag)^{ij} \right) \chi_a^j - \frac{\mathrm{tr}\,V^{\dag}V}{g^2/N} \right] \nonumber \\
    &= \int \mathcal{D} \left[ \bar{\psi}, \psi, \bar{\chi}, \chi \right] \exp \bigg[ - \bar{\psi}_a^{i}\,( G_0^{-1} )^{ij} \psi_a^j \nonumber \\ 
    &\qquad\qquad - \bar{\chi}_a^{i}\,( (G_0^{-1})^{\dag} )^{ij} \chi_a^j + \frac{g^2}{N} \bar{\psi}^{i}_a \psi^j_a \bar{\chi}^j_b \chi^i_b \bigg]
\end{align}
with the Green's function without disorder:
\begin{equation}
    G_0 \coloneqq \left( E - H_0 \right)^{-1}.
\end{equation}

Importantly, although we begin with quadratic systems, the disorder average yields the quartic term $\propto \bar{\psi}^{i}_a \psi^j_a \bar{\chi}^j_b \chi^i_b$ in the action.
Using the Hubbard-Stratonovich transformation, we introduce an $n \times n$ auxiliary matrix field $Q$ and decouple this quartic term into quadratic terms:
\begin{align}
    &\exp \left[ \frac{g^2}{N} \bar{\psi}_a^i \psi_a^j \bar{\chi}_b^j \chi_b^i \right] \nonumber \\
    &= \int \mathcal{D}Q~\exp \left[ - \frac{N}{g^2} Q_{ab}^{*} Q_{ab} + \bar{\psi}_{a}^i Q_{ab} \chi_b^i + \psi_{a}^i Q_{ab}^{*} \bar{\chi}_b^i \right].
\end{align}
Now that the action is bilinear in the Grassmann variables, the integral is performed explicitly with the Gaussian integral.
The replica partition function further reduces to
\begin{align}
    Z_n &= \int \mathcal{D}Q~e^{-(N/g^2)\,\mathrm{tr}\,Q^{\dag}Q} \int \mathcal{D} \left[ \bar{\psi}, \psi, \bar{\chi}, \chi \right] \nonumber \\
    &\qquad\quad \exp \left[ - \begin{pmatrix}
        \bar{\psi} & \bar{\chi}
    \end{pmatrix} \begin{pmatrix}
        G_0^{-1} & - Q \\
        Q^{\dag} & (G_0^{-1})^{\dag}
    \end{pmatrix} \begin{pmatrix}
        \psi \\ \chi
    \end{pmatrix} \right] \nonumber \\
    &= \int \mathcal{D}Q~e^{-(N/g^2)\,\mathrm{tr}\,Q^{\dag}Q} \left[ \det \begin{pmatrix}
        G_0^{-1} & - Q \\
        Q^{\dag} & (G_0^{-1})^{\dag}
    \end{pmatrix} \right]^N \nonumber \\
    &\eqqcolon \int \mathcal{D}Q~e^{-N S_n [Q]}
\end{align}
with the effective action
\begin{equation}
    S_n [Q] \coloneqq \frac{\mathrm{tr}\,Q^{\dag}Q}{g^2} - \mathrm{tr} \log \begin{pmatrix}
        G_0^{-1} & - Q \\
        Q^{\dag} & (G_0^{-1})^{\dag}
    \end{pmatrix}.
        \label{eq: SnQ}
\end{equation}
Through the Hubbard-Stratonovich transformation, the matrix integral over large $N \times N$ random systems $H$ changes into that over the ``small" $n \times n$ matrix $Q$.
In the large-$N$ limit $N \to \infty$, the $Q$-integral is dominated by the saddle point of the effective action $S_n [Q]$.
Depending on symmetry of the microscopic non-Hermitian Hamiltonians $H$, the saddle points of $Q$ form different target manifolds, which further characterize the universality classes of spectral correlations and possible Anderson transitions.
For example, in the absence of symmetry (i.e., class A), the target manifold of $Q$ at the saddle point is the unitary group \Cone~\cite{Nishigaki-02}.

%%%%%%
\subsection{Classification}
    \label{subsec: classification}

To study symmetry of the nonlinear sigma models in Eq.~(\ref{eq: SnQ}), the Hermitization method plays a crucial role.
For a given single-particle non-Hermitian Hamiltonian $H$ and complex energy $E \in \mathbb{C}$, we construct a single-particle Hermitian Hamiltonian $\tilde{H}_E$ by
\begin{equation}
    \tilde{H}_E \coloneqq \begin{pmatrix}   
        0 & H-E \\
        H^{\dag} - E^{*} & 0 
    \end{pmatrix}.
        \label{eq: Hermitization}
\end{equation}
By construction, the Hermitized Hamiltonian $\tilde{H}_E$ respects additional chiral symmetry, 
\begin{equation}
    \Gamma \tilde{H}_E \Gamma^{-1} = - \tilde{H}_E, \quad \Gamma \coloneqq \begin{pmatrix}
        1 & 0 \\
        0 & -1
    \end{pmatrix}.
\end{equation}
This Hermitization method was introduced to calculate the spectral statistics of non-Hermitian random matrices~\cite{Girko-85, Feinberg-97}.

Notably, the replica partition function $Z_n \left( E, E^{*} \right)$ in Eq.~(\ref{eq: replica partition function}) is given as
\begin{equation}
    Z_n \left( E, E^{*} \right) = \left( -1 \right)^n \Braket{[ \det \tilde{H}_E ]^n}.
        \label{eq: ZnE - Hermitization}
\end{equation}
Thus, $Z_n \left( E, E^{*} \right)$ for non-Hermitian systems $H$ coincides with that for Hermitized systems $\tilde{H}_E$ with chiral symmetry.
Consequently, the target manifolds of $Q$ in the nonlinear sigma models $S_n [Q]$ reduce to those for Hermitian systems with additional chiral symmetry.
In this manner, using the classification for Hermitian systems (see Table~\ref{tab: Hermitian})~\cite{Evers-review}, we develop the classification of nonlinear sigma models for non-Hermitian disordered systems in all the 38 symmetry classes, as summarized in Tables~\ref{tab: complex AZ}-\ref{tab: real AZ + SLS}.
Owing to additional chiral symmetry, the target manifolds of nonlinear sigma models differ from those in the Hermitian case, and concomitantly, the universality classes of the Anderson transitions change.
For example, as also mentioned before, the saddle-point target manifold of $Q$ in the absence of symmetry (i.e., class A) forms the unitary group \Cone, which coincides with the target manifold for Hermitian systems with chiral symmetry (i.e., class AIII).

Notably, Ref.~\cite{Luo-22R} showed that the critical behavior of the length scale in non-Hermitian disordered systems coincides with that in the corresponding Hermitian systems with additional chiral symmetry, using microscopic lattice models.
Our work corroborates this correspondence in the field-theoretical perspective;
it is a consequence of the duality of the replica partition functions between non-Hermitian systems and Hermitized systems with chiral symmetry.
One of the remarkable consequences of Ref.~\cite{Luo-22R} is the superuniversality of Anderson transitions in non-Hermitian disordered systems.
For example, in Ref.~\cite{Luo-22R}, the critical exponents of the localization length are shown to be common between two-dimensional non-Hermitian disordered systems in classes DIII and CII$^{\dag}$ and reduce to that in Hermitian systems in class AII.
Consistently, the target manifold of the underlying nonlinear sigma models in these symmetry classes is \Rfour~(see Tables~\ref{tab: real AZ} and \ref{tab: real AZ-dag}), implying the same universality class.
Such duality of non-Hermitian disordered systems in different symmetry classes is found in a number of other symmetry classes, as summarized in Tables~\ref{tab: complex AZ}-\ref{tab: real AZ + SLS}.
As a direct consequence, we only have the ten different types of target manifolds of nonlinear sigma models even in the non-Hermitian case.

It is also notable that topological terms can be added to the nonlinear sigma models in a similar manner to the topological field theory description in Ref.~\cite{KSR-21}.
The classification of possible topological terms coincides with that of non-Hermitian topology with respect to point gaps~\cite{Gong-18, KSUS-19}, which is also listed in Tables~\ref{tab: complex AZ}-\ref{tab: real AZ + SLS} for one, two, and three spatial dimensions.
In fact, the classification of point-gap topology for non-Hermitian systems $H$ coincides with the topological classification of Hermitized systems $\tilde{H}_E$ in Eq.~(\ref{eq: Hermitization}).
For example, non-Hermitian disordered systems with no symmetry are assigned to topological terms for odd spatial dimensions.
In Sec.~\ref{sec: Hatano-Nelson}, we show that such a topological term is the origin of the Anderson transitions in nonreciprocal disordered systems in one dimension, including the Hatano-Nelson model~\cite{Hatano-Nelson-96, *Hatano-Nelson-97}.
In Sec.~\ref{sec: others}, we discuss the relevance of the topological terms to non-Hermitian disordered systems in other symmetry classes and/or in higher dimensions.
Furthermore, $E$ should be chosen to respect symmetry to realize the symmetry-preserving saddle-point target manifolds.
If $E$ breaks relevant symmetry, the target manifolds reduce to those without symmetry.
As an example, we discuss non-Hermitian random matrices in class D in Sec.~\ref{sec: RMT}.

%%%%%%
\subsection{Comparison with the Hermitian regime}
    \label{subsec: Hermitian}

As discussed above, non-Hermiticity changes the target manifolds of nonlinear sigma models and concomitant universality classes of the Anderson transitions.
Here, we show that this difference is due to the distinct formulations of the Green's functions and partition functions.
For Hermitian systems $H$, the spectra are real-valued, and hence the Green's function in Eq.~(\ref{eq: Green}) is given as
\begin{equation}
    G \left( E \right) = \frac{1}{N} \frac{\partial}{\partial E} \braket{\log \det \left( E-H \right)},
\end{equation}
where $E$ is chosen to be real.
As a result, instead of Eq.~(\ref{eq: replica partition function}), the replica partition function for Hermitian systems is chosen as
\begin{equation}
    Z_n \left( E \right) \coloneqq \braket{\left[ \det \left( E-H \right) \right]^n},
        \label{eq: ZnE - Hermitian}
\end{equation}
from which the Green's function reads
\begin{equation}
    G \left( E \right) = \lim_{n\to 0} \frac{1}{n} \frac{\partial}{\partial E} Z_n \left( E \right).
\end{equation}
Equation~(\ref{eq: ZnE - Hermitian}) is equivalent to Eq.~(\ref{eq: ZnE - Hermitization}) for $\tilde{H}_E$ and zero energy.
Additionally, from the identity 
\begin{equation}
    \lim_{\varepsilon \to 0^{+}} \mathrm{Im}\,\frac{1}{E+\ii \varepsilon} = - \pi \delta \left( E \right)
        \label{eq: delta - 1D}
\end{equation}
for $E \in \mathbb{R}$, the density of states is given as 
\begin{equation}
    \rho \left( E \right) = - \frac{1}{\pi} \lim_{\varepsilon \to 0^{+}}\mathrm{Im}\,G \left( E+\ii \varepsilon \right).
        \label{eq: rho - 1D}
\end{equation}

Importantly, the above formulas are valid only if the spectra of $H$ are real-valued and form one-dimensional segments in the complex plane.
Specifically, the analytic continuation in Eqs.~(\ref{eq: delta - 1D}) and (\ref{eq: rho - 1D}) works since singularities of the Green's functions are distributed only on the real axis.
Conversely, it no longer works when singularities of the Green's functions are extended to two-dimensional regions in the complex plane because of non-Hermiticity of $H$; 
instead, the different formulas in Eqs.~(\ref{eq: delta - 2D}) and (\ref{eq: rho - 2D}) should be used.
Such a spectral difference between Hermitian and non-Hermitian systems forces us to employ distinct choices of the replica partition functions in Eqs.~(\ref{eq: replica partition function}) and~(\ref{eq: ZnE - Hermitian}), further leading to the different target manifolds of nonlinear sigma models.

It is also notable that even non-Hermitian disordered systems can host almost real-valued spectra as long as non-Hermiticity is sufficiently weak~\cite{Fyodorov-Sommers-review}.
In such a weakly non-Hermitian regime, the replica partition function in Eq.~(\ref{eq: ZnE - Hermitian}) is applicable, and the target manifolds of nonlinear sigma models can reduce to the Hermitian ones.
In this work, by contrast, we focus on generic non-Hermitian disordered systems in which Hermiticity and non-Hermiticity are comparable with each other.
While the weakly non-Hermitian regime should correspond to classifying spaces based on line gaps, the generic non-Hermitian regime should correspond to those based on point gaps, the combination of which constitutes the 38-fold classification of non-Hermitian systems~\cite{KSUS-19}.

%%%%% RMT %%%%%
\section{Non-Hermitian random matrices}
    \label{sec: RMT}

As an illustrative example, we develop a nonlinear sigma model description of non-Hermitian random matrices with symmetry (i.e., non-Hermitian disordered systems in zero dimension).
Specifically, we study $N \times N$ non-Hermitian random matrices in class D, respecting particle-hole symmetry
\begin{equation}
    \mathcal{C} H^{T} \mathcal{C}^{-1} = -H
        \label{eq: PHS}
\end{equation}
with a unitary matrix $\mathcal{C}$ satisfying $\mathcal{C}\mathcal{C}^{*} = +1$.
Without loss of generality, we choose $\mathcal{C}$ to be the identity matrix, for which $H$ is a random antisymmetric matrix (i.e., $H^{T} = - H$).
We further assume the Gaussian distribution $\propto e^{-(1/2)\,\mathrm{tr}\,H^{\dag}H}$.

In Sec.~\ref{subsec: D-derivation}, we explicitly derive an effective nonlinear sigma model from non-Hermitian random matrices in class D.
In Sec.~\ref{subsec: D-saddle}, we perform the saddle-point analysis for the large-$N$ limit and show that the target manifold of the nonlinear sigma model for class D is the orthogonal group \Rthree, consistent with Table~\ref{tab: real AZ}.
In Sec.~\ref{subsec: D-CP}, we present detailed calculations of characteristic polynomials.
In Sec.~\ref{subsec: D-duality}, we discuss the duality of non-Hermitian random matrices as a unique feature to non-Hermitian disordered systems.
In Sec.~\ref{subsec: NH RMT topology}, we consider the relevance of topological terms to non-Hermitian random matrices.
In Appendix~\ref{asec: NH RMT}, we describe the details of our discussion.
In particular, we discuss the characteristic polynomials in other cases, such as class C.

%%%%%%%%%%%%
\subsection{Derivation}
    \label{subsec: D-derivation}

The replica partition function in Eq.~(\ref{eq: replica partition function}) reads
\begin{align}
    Z_n &= \int \mathcal{D} \left[ \bar{\psi}, \psi, \bar{\chi}, \chi, H \right] \exp \left[ - \bar{\psi}_a^i \left( E\delta^{ij} -H^{ij} \right) \psi_a^j \right. \nonumber \\
    &\qquad\quad \left. - \bar{\chi}_a^i \left( E^{*} \delta^{ij} - (H^{\dag})^{ij}  \right) \chi_a^j - \frac{\mathrm{tr}\,H^{\dag}H}{2} \right].
\end{align}
Owing to particle-hole symmetry, the diagonal elements $H^{ii}$ vanish;
the off-diagonal elements $H^{ij}$ ($i < j$) are not independent of the other off-diagonal elements $H^{ij}$ ($i > j$) but determined by $H^{ij} = - H^{ji}$.
The Gaussian integral over the matrix elements $H^{ij}$ ($i > j$) yields
\begin{align}
    &\int \mathcal{D}H \exp \left[ \bar{\psi}_a^i H^{ij} \psi_a^j + \bar{\chi}_a^i (H^{\dag})^{ij} \chi_a^j - \frac{\mathrm{tr}\,H^{\dag}H}{2} \right] \nonumber \\
    &= \int \mathcal{D}H \exp \left[ \sum_{i>j} \left( H^{ij} \Psi^{ij} - (H^{ij})^{*} X^{ij} - (H^{ij})^{*} H^{ij} \right) \right] \nonumber \\
    %&= \int \mathcal{D}H \exp \left[ -\sum_{i>j} \left( \left( (H^{ij})^{*} - \Psi^{ij} \right) \left( H^{ij} + X^{ij} \right) + \Psi^{ij} X^{ij} \right) \right] \nonumber \\
    &= \exp \left[ - \sum_{i>j} \Psi^{ij} X^{ij} \right] \nonumber \\
    &= \exp \left[ \bar{\psi}_a^i \psi_a^j \bar{\chi}_b^j \chi_b^i + \bar{\psi}_a^i \psi_a^j \chi_b^j \bar{\chi}_b^i \right]
\end{align}
with $\Psi^{ij} \coloneqq \bar{\psi}_a^i \psi_a^j + \psi_a^i \bar{\psi}_a^j$ and $X^{ij} \coloneqq \bar{\chi}_a^i \chi_a^j + \chi_a^i \bar{\chi}_a^j$.
To decouple the quartic terms into quadratic terms, we use the Hubbard-Stratonovich transformation,
\begin{align}
    &\exp \left[ \bar{\psi}_a^i \psi_a^j \bar{\chi}_b^j \chi_b^i\right] \nonumber \\
    &\quad = \int\mathcal{D}P \exp \left[ - P^{*}_{ab} P_{ab} + \bar{\psi}_a^i P_{ab} \chi_b^i + \psi_a^i P^{*}_{ab} \bar{\chi}_b^i\right], \\
    &\exp \left[ \bar{\psi}_a^i \psi_a^j \chi_b^j \bar{\chi}_b^i\right] \nonumber \\
    &\quad = \int\mathcal{D}R \exp \left[ - R^{*}_{ab} R_{ab} + \bar{\psi}_a^i R_{ab} \bar{\chi}_b^i + \psi_a^i R^{*}_{ab} \chi_b^i \right]
\end{align}
with the $n \times n$ auxiliary matrix fields $P, R \in \mathbb{C}^{n\times n}$.
Using $P$ and $R$, we reduce the replica partition function $Z_n$ to 
\begin{align}
    Z_n &= \int \mathcal{D}\left[ P, R\right]e^{-\mathrm{tr}\,[P^{\dag}P + R^{\dag}R]} \int \mathcal{D} \left[ \bar{\psi}, \psi, \bar{\chi}, \chi \right] \nonumber \\
    &\exp \left[ - \frac{1}{2} \begin{pmatrix}
        \bar{\psi} & \psi & \bar{\chi} & \chi
    \end{pmatrix} \begin{pmatrix}
        0 & E & -R & -P \\
        -E & 0 & -P^* & -R^* \\
        R^T & P^\dag & 0 & E^* \\
        P^T & R^\dag & -E^* & 0\\
    \end{pmatrix} \begin{pmatrix}
        \bar{\psi} \\ \psi \\ \bar{\chi} \\ \chi
    \end{pmatrix} \right] \nonumber \\
    \label{eq: class D HS transform}
    &= \int \mathcal{D}Q~e^{-(1/2)\,\mathrm{tr}\,Q^{\dag}Q} \left[ \det \begin{pmatrix}
        \ii E\sigma_y & -Q \\
        Q^T & \ii E^{*} \sigma_y
    \end{pmatrix} \right]^{N/2}
\end{align}
with the $2n \times 2n$ matrix field
\begin{equation}
    Q \coloneqq \begin{pmatrix}
        R & P \\
        P^* & R^*
    \end{pmatrix}.
\end{equation}
Notably, by construction, $Q$ is subject to the constraint
\begin{equation}
    \sigma_x Q^{*} \sigma_x = Q.
        \label{eq: Q - reality}
\end{equation}
Then, the effective action $S_n [Q]$, defined by $Z_n \eqqcolon \int \mathcal{D}Q~e^{-(N/2)\,S_n [Q]}$, is given as
\begin{align}
    S_n [Q] &= \frac{1}{N}\mathrm{tr}\,Q^{\dag}Q - \log \det \begin{pmatrix}
        \ii E\sigma_y & -Q \\
        Q^T & \ii E^{*} \sigma_y
    \end{pmatrix} \nonumber \\
    &= \mathrm{tr} \left[ \frac{1}{N} Q^{\dag} Q - \log \left( \left| E \right|^2 + \sigma_y Q^T \sigma_y Q \right) \right].
\end{align}

%%%%%%
\subsection{Saddle-point target manifold}
    \label{subsec: D-saddle}

In the large-$N$ limit, the partition function $Z_n$ is dominated by the saddle point of the effective action $S_n [Q]$ (i.e., $\delta S_n/\delta Q = 0$).
We focus on the behavior around the spectral origin $\left| E \right| \ll \sqrt{N}$, which accompanies the universal spectral statistics due to particle-hole symmetry.
In this microscopic regime, the saddle point of $S_n [Q]$ reads, 
\begin{equation}
    Q^{\dag}Q = N.
        \label{eq: Q - unitarity}
\end{equation}
On the basis of the reality condition in Eq.~(\ref{eq: Q - reality}) and the unitarity condition in Eq.~(\ref{eq: Q - unitarity}), $Q$ at the saddle point is generally expressed as
\begin{equation}
    \label{eq: Q - saddle point}
    Q = \sqrt{N} \sigma_x^{1/2} O \sigma_x^{1/2}, \quad O \in \mathrm{O} \left( 2n \right).
\end{equation}
Consequently, the saddle-point target manifold of the nonlinear sigma model for class D is the orthogonal group $\mathrm{O} \left( 2n \right)$.
This coincides with that for Hermitian disordered systems in class DIII (see Table~\ref{tab: Hermitian}), consistent with our discussion in Sec.~\ref{subsec: classification}.
The universal level statistics around the spectral origin for non-Hermitian random matrices in class D were studied for microscopic models both analytically and numerically~\cite{Splittorff-04, Akemann-09, GarciaGarcia-22PRX, Xiao-24}.
For example, particle-hole symmetry leads to a larger density of states around the spectral origin in comparison with the circular law.
%It should be significant to thoroughly study such universal spectral statistics from the field-theoretical perspective.

%%%%%%
\subsection{Characteristic polynomial}
    \label{subsec: D-CP}

As an application of our replica formalism, we study the characteristic polynomials for non-Hermitian random matrices in class D in more detail.
While the correct density of states seems challenging to derive accurately, we illustrate the computation of relevant physical observables within our replica nonlinear sigma model framework.
This discussion further clarifies how symmetry dictates different universality classes of non-Hermitian disordered systems.
As discussed above, in the large-$N$ limit, the integration over $Q$ 
in Eq.~\eqref{eq: class D HS transform} 
can be restricted to the saddle-point manifold given by Eq.~\eqref{eq: Q - saddle point},
\begin{equation}
    \label{eq:class d saddle point}
    Z_n(E) \simeq \int {\cal D}{Q} \left[\det\begin{pmatrix}
        \ii \left| E \right|\sigma_y & Q \\
        -Q^T & \ii \left| E \right| \sigma_y
    \end{pmatrix}\right]^{N/2},
\end{equation}
where ${\cal D}Q$ represents the Haar measure of 
$\mathrm{O} \left( 2n \right)$.
The integral over $Q$ is divided into the two connected components of 
$\mathrm{O} \left( 2n \right)$:
$\mathrm{SO} \left( 2n \right)$ and its complement,
$\mathrm{O}\left(2n\right)
\setminus \mathrm{SO}\left(2n\right)$.
Accordingly, the partition function is given by 
the sum of the contributions from each sector, $Z_n(E)=(Z_{n,+}(E)+Z_{n,-}(E))/2$.

We simplify the matrix integral by using the symmetry under $Q \mapsto UQV$ for $U,V\in \mathrm{OSp}\left(2n\right) \cong \mathrm{U}(n)$, where $\mathrm{OSp}(2n)$ is a subgroup of $\mathrm{O}(2n)$ whose elements commute with $\sigma_y$.
We thus reduce the matrix integral to the multiple integrals over $\lambda_i$ defined below.
Specifically, we use the following decomposition of the $Q$ matrix~\cite{Edelman-23}.
If $n = 2k+1$ is odd, for any ${Q}\in\operatorname{SO}(2n)$, there exist unique angles (up to ordering) $\theta_i\in[0,\pi/2)$ for $i = 1,\cdots,k$ and $U,V\in \operatorname{OSp}(2n)$ such that
\begin{equation}
\label{eq:rho_upv_odd}
        {Q} = U P V, \quad P \coloneqq \begin{pmatrix}
            \rho & \\ & \rho^T
        \end{pmatrix}
\end{equation}
with
\begin{equation}
        \rho \coloneqq \begin{pmatrix}
            1 & & & \\ & \rho_{\theta_1} \\ & & \ddots \\ & & & \rho_{\theta_{k}}
        \end{pmatrix}.
\end{equation}
Additionally, for $n=2k+1$ and $Q\in \mathrm{O}(2n)\setminus \mathrm{SO}(2n)$, a similar decomposition holds if we use
\begin{equation}
    P \coloneqq \begin{pmatrix}
        \rho & \\ & -\rho^T
    \end{pmatrix}.
\end{equation}
On the other hand, if $n = 2k$ is even, for any ${Q}\in\operatorname{SO}(2n)$, there exist unique (up to ordering) angles
$\theta_i\in[0,\pi/2)$ for $i = 1,\cdots,k$ and $U,V\in \operatorname{OSp}(n)$ such that
\begin{equation} \label{eq:rho_upv_even}
  {Q} = U P V, \quad P \coloneqq \begin{pmatrix}
            \rho & \\ & \rho^T
        \end{pmatrix},
\end{equation}
    with
\begin{equation}
   \rho \coloneqq \begin{pmatrix}
   \rho_{\theta_1} \\ & \ddots \\ & & \rho_{\theta_{k}}
   \end{pmatrix},
   \quad
  \rho_{\theta} \coloneqq
  \begin{pmatrix}
  \cos\theta &\sin\theta \\
  -\sin\theta & \cos\theta
  \end{pmatrix}.
\end{equation}

Using these decompositions, 
we express the partition functions
$Z_{2k+1,\pm}(E)$ and $Z_{2k,+}(E)$
as integrals over
$\theta_i$'s.
By further defining 
$\lambda_i \coloneqq \cos 2\theta_i$,
the integral on $\mathrm{SO}(2n)$ for even $n=2k$ 
is written as 
\begin{align}
    Z_{2k,+}(E) &= \int_{[-1,1]^k} \left( \prod_{\ell=1}^k d{\lambda_\ell}\right)  |\Delta(\lambda)|^4
    \nonumber \\
    &\phantom{{}={}} 
    \times \prod_{\ell=1}^k \left( \left|E \right|^4 - 2 \left|E \right|^2 N \lambda_\ell + N^2 \right)^{N},
\end{align}
where $\Delta(\lambda)$ is
the Vandermonde determinant,
$\Delta(\lambda) \coloneqq \prod^k_{i<j} (\lambda_i-\lambda_j)$.
For large $N$, the integrand can be approximated as (up to a constant factor), 
\begin{align}
    Z_{n,+}(E) &\simeq e^{2k|E|^2} \int_{[0,1]^k} \left( \prod_{\ell=1}^k d{\lambda_\ell}\right)  |\Delta(\lambda)|^4
    \nonumber \\
    &\phantom{{}={}} \quad 
    \times \exp\left(- 4 \left| E \right|^2\sum_{\ell=1}^k \lambda_\ell\right).
        \label{eq: D k-fold integral}
\end{align}

To express $Z_{2k+1,\pm}(E)$ 
and $Z_{2k,\pm}(E)$ in a 
unified manner,
let us define
the following integrals,
\begin{align}
\mathcal{I}^{(\alpha,\beta,\gamma)}_k(t) &\coloneqq 
\int_{[0,1]^k} 
 \left( \prod_{\ell=1}^k d{\lambda_\ell}~\lambda_\ell^{\alpha-1} (1-\lambda_\ell)^{\beta-1} \right) |\Delta(\lambda)|^{2\gamma}
\nonumber \\
    &
    \qquad \qquad \qquad
    \times
\exp\left[t\sum_{\ell=1}^k \lambda_\ell\right], \\
    {I}^{(\alpha,\beta,\gamma)}_k(t) &\coloneqq \frac{\mathcal{I}^{(\alpha,\beta,\gamma)}_k(t)}{\mathcal{I}^{(\alpha,\beta,\gamma)}_k(0)}.
\end{align}
Then, for $n=2k+1$,
the above decompositions lead to 
\begin{align}
    Z_n(E) &= \frac{1}{2} e^{(n-2)|E|^2} {I}^{(1,3,2)}_{(n-1)/2}(-4|E|^2) \nonumber \\
    &\phantom{{}={}} \qquad \quad + \frac{1}{2}e^{n |E|^2} {I}^{(3,1,2)}_{(n-1)/2}(-4|E|^2), \label{eq: class d saddle pt odd}
\end{align}
where the right-hand side is normalized by $Z_n(0) = 1$.
Similarly, 
for $n=2k$,
$Z_{n,+}(E)$ is expressed as 
\begin{align}
    Z_{n,+}(E) = e^{n|E|^2} {I}^{(1,1,2)}_{n/2}(-4|E|^2).
        \label{eq: D n+}
\end{align}
While we can derive the analytical results for $Z_{2k,+}(E)$ and $Z_{2k+1,\pm}(E)$,
a similar decomposition of $Q$ is unknown for $Z_{2k,-}(E)$.
Nevertheless, based on numerical integrals over $Q$ for small $n$, 
we conjecture
\begin{align}
Z_{n,-}(E) = e^{(n-2)|E|^2} {I}^{(3,3,2)}_{n/2-1}(-4|E|^2).
        \label{eq: D n- conjecture}
\end{align}
In summary, for $n=2k$, we have
\begin{align}
    Z_n(E) &= \frac{1}{2} e^{n|E|^2} {I}^{(1,1,2)}_{n/2}(-4|E|^2)
    \nonumber 
    \\
    &\phantom{{}={}} \quad +\frac{1}{2} e^{(n-2)|E|^2} {I}^{(3,3,2)}_{n/2-1}(-4|E|^2). \label{eq: class d saddle pt even}
\end{align}
While Eq.~(\ref{eq: D n+}) is derived analytically, Eq.~(\ref{eq: D n- conjecture}) is a conjecture.
Consequently, Eq.~(\ref{eq: class d saddle pt even}) is partially a conjecture.
However, we confirm that Eq.~(\ref{eq: class d saddle pt even}) coincides with numerically obtained characteristic polynomials (see Table~\ref{tab: comparison saddle point vs small N} in Appendix~\ref{asec: NH RMT} for details).

We express the integral $I^{(\alpha,\beta,\gamma)}_k(t)$ in terms of confluent hypergeometric functions associated with the Jack polynomials as~\cite{Yan1992, Kaneko1993}
\begin{equation}
    \label{eq: hypergeometric integral}
    \frac{I^{(\alpha,\beta,\gamma)}_k(t)}{I^{(\alpha,\beta,\gamma)}_k(0)} = \confluenthypergeometricjack{1/\gamma}\left(\begin{array}{c}
        \alpha + \gamma \left( k-1 \right) \\ \alpha+\beta+2\gamma \left( k-1 \right)
    \end{array};t^{\oplus k}\right).
\end{equation}
Here, $\confluenthypergeometricjack{1/\gamma}$ is given by 
\begin{equation}
\confluenthypergeometricjack{1/\gamma}\left(\begin{array}{c}
        a \\ c
    \end{array};t^{\oplus k}\right)
    \coloneqq \sum_{m=0}^\infty \sum_{\mu\in\mathcal{P}(m)} \frac{(a)_\mu}{(b)_\mu} \frac{C_\mu^{(1/\gamma)}
    (t, k)}{m!},
\end{equation}
where $\mathcal{P}(m)$ is the set of possible partitions of a natural number $m \in \mathbb{N}$, $(a)_\mu$ is the generalized Pochhammer symbol, and $C^{(1/\gamma)}_\mu(t,k)$ 
is defined in 
terms of the Jack polynomial
$C^{(1/\gamma)}_{\mu}(t_1,\cdots, t_k)$
as 
$
C^{(1/\gamma)}_\mu(t, k) 
\coloneqq
C^{(1/\gamma)}_\mu(t^{\oplus k}) = C^{(1/\gamma)}_\mu(\underbrace{t,t,\cdots,t}_{k\text{ times}})$.
We also provide details on hypergeometric functions in Appendix~\ref{asec: hypergeometric}.

A priori, $C^{(1/\gamma)}_{\mu}(t,k)$
is defined only for $k \in \mathbb{N}$. 
However, it can be extended to general $k\in \mathbb{R}$ by
\begin{equation}
    C_\mu^{(1/\gamma)}(t, k) = (\gamma k)_\mu  \frac{C_\mu^{(1/\gamma)}(t, \operatorname{len}(\mu))}{(\gamma \cdot \operatorname{len}(\mu))_\mu},
\end{equation}
where $\operatorname{len}(\mu)$ denotes the length of the partition $\mu \in \mathcal{P} \left( m \right)$, i.e., the number of nonzero elements in $\mu$.
Through this analytic continuation, we finally obtain
\begin{align}
    &Z_n \left( E \right) = 1 + \frac{n}{4n-2} \left| E \right|^4 
    \nonumber \\
    &\phantom{{}={}} \quad
    +\frac{n \left(2 n^2-n-2\right)}{8 \left(8 n^3-12 n^2-2 n+3\right)} \left| E \right|^8 + \cdots.
        \label{eq: ZnE - replica}
\end{align}
For some specific choices of $n$ and $N$, we calculate $Z_n \left( E \right)$ analytically or numerically, confirming this result;
see Table~\ref{tab: comparison saddle point vs small N} in Appendix~\ref{asec: NH RMT} for details.
Since our result from the nonlinear sigma model is valid for large $N$, 
the coincidence with the finite-$N$ result is not exact. 
Nevertheless, both calculations are consistent for increasing $N$.

%Although Eq.~(\ref{eq: ZnE - replica}) gives correct $Z_n$ for integer $n$, it fails for fractional $n$ due to a branch cut of $Z_n$ at the origin.
Equation~(\ref{eq: ZnE - replica}), combined with Eqs.~(\ref{eq: DOS - potential}) and (\ref{eq: Phi - ZnE}), leads to the density of states
\begin{align}
    \rho \left( E \right) &= \frac{1}{\pi} \frac{\partial^2}{\partial E \partial E^{*}} \left. \frac{\partial Z_n \left( E \right)}{\partial n} \right|_{n=0} \nonumber \\ 
    &= - \frac{\sinh\,( 2 \left| E \right|^2 )}{\pi}.
        \label{eq: DOS-D-incorrect}
\end{align}
The obtained density of states vanishes at $E=0$ and becomes negative for $\left| E \right| > 0$, which is definitely incorrect.
We expect that this incorrect result should be compensated with a missing contribution that cannot be readily captured by the replica formalism, such as $\cosh\,( 2 \left| E\right|^2 )/\pi$.
Still, it should partially account for the peak of the density of states at the spectral origin.

For positive integer values of $n$, the characteristic polynomial $Z_n$ can be expressed as $k$-fold integrals within the framework of nonlinear sigma models [see, for example, Eq.~(\ref{eq: D k-fold integral})].
These $k$-fold integrals can subsequently be evaluated by hypergeometric functions, yielding correct $Z_n$ for positive integers $n$.
For noninteger values of $n$, on the other hand, $k$ also becomes nonintegers, making the direct evaluation of the $k$-fold integrals difficult.
In general, $Z_n$ includes terms involving fractional powers of $\left| E \right|$.
Consequently, $Z_n$ for arbitrary nonintegers $n$ cannot be uniquely determined solely by the straightforward analytic continuation of $Z_n$ obtained for integers $n$.
Our results imply that class D belongs to this category.
Nevertheless, the correct density of states can be accurately obtained in other symmetry classes, such as classes A, AI, and AII~\cite{Nishigaki-02}.

%%%%%%
\subsection{Duality between different symmetry classes}
    \label{subsec: D-duality}

It is also notable that non-Hermitian disordered systems in different symmetry classes can be described by the same nonlinear sigma models, as also discussed in Sec.~\ref{subsec: classification}.
Specifically, non-Hermitian disordered systems in class D share the same nonlinear sigma model with those in class AII$^{\dag}$, which respect time-reversal symmetry$^{\dag}$ as a defining feature,
\begin{equation}
    \mathcal{T} H^{T} \mathcal{T}^{-1} = H,
        \label{eq: TRSdag}
\end{equation}
with a unitary matrix $\mathcal{T}$ satisfying $\mathcal{T}\mathcal{T}^{*} = -1$~\cite{KSUS-19}.
In fact, for a given non-Hermitian random matrix $H$ that respects time-reversal symmetry$^{\dag}$, another non-Hermitian random matrix $H\mathcal{T}$ respects particle-hole symmetry in Eq.~(\ref{eq: PHS}) with $\mathcal{C} = 1$,
\begin{equation}
    \left( H\mathcal{T} \right)^T = \mathcal{T}^T H^T = - \mathcal{T}H^T = - H \mathcal{T},
\end{equation}
where $\mathcal{T}^T = - \mathcal{T}$ is used.
Consequently, classes AII$^{\dag}$ and D share the same partition function,
\begin{align}
    &Z_n^{{\rm AII}^{\dag}} \left( E, E^{*} \right) = \Braket{\left( -1 \right)^n \left[ \det \begin{pmatrix}
        0 & H-E \\
        \left( H-E \right)^{\dag} & 0
    \end{pmatrix}\right]^n} \nonumber \\
    &\qquad \propto \Braket{\left( -1 \right)^n \left[ \det \begin{pmatrix}
        0 & \left( H-E \right) \mathcal{T} \\
        \left[ \left( H-E \right) \mathcal{T} \right]^{\dag} & 0
    \end{pmatrix}\right]^n} \nonumber \\ 
    &\qquad = Z_n^{{\rm D}} \left( 0, 0 \right).
\end{align}
This further results in the same target manifold \Rthree~of the nonlinear sigma model, as well as the same universality class of the Anderson transitions~\cite{Luo-22R}, consistent with our classification (see Tables~\ref{tab: real AZ} and \ref{tab: real AZ-dag}).
It should be noted that the partition function $Z_n^{{\rm D}} \left( 0, 0 \right)$ for class D manifests the duality only around $E = E^{*} = 0$; 
the level statistics around the spectral origin for class D and those within the spectral bulk for class AII$^{\dag}$ are described by the same effective field theory.

Additionally, in a similar manner to our analysis in Sec.~\ref{subsec: D-CP}, the partition function for class C in the large-$N$ limit is obtained as
\begin{equation}
\label{class c result}
    Z^{\mathrm{C}}_n \left( E, E^{*} \right) = e^{n |E|^2} {I}^{(1,1,1/2)}_n\,(-2 \left| E \right|^2 ),
\end{equation}
satisfying 
\begin{equation}
\label{eq: class d and c duality}
Z^{\mathrm{C}}_n \left( E, E^{*} \right) = Z^{\mathrm{D}}_{-n} \left( E, E^{*} \right)
\end{equation}
for $n \in \mathbb{N}$ (see Appendix~\ref{asec: NH RMT} for details).
This is a manifestation of the duality between $\mathrm{Sp} \left( n \right)$ and $\mathrm{O} \left( -2n \right)$~\cite{Wegner-1981}.
From this relationship, if the density of states in class D displays a peak at the spectral origin, that in class C should conversely exhibit a dip, which is consistent with the numerical results~\cite{GarciaGarcia-22PRX, Xiao-24}.

%%%%%%
\subsection{Topology}
    \label{subsec: NH RMT topology}

As discussed previously, nonlinear sigma models can incorporate topological terms.
This is also the case for non-Hermitian random matrices, in which zero-dimensional topological terms are relevant.
While such zero-dimensional topological terms are not explicitly listed in Tables~\ref{tab: complex AZ}-\ref{tab: real AZ + SLS}, they can be identified in Tables~III-IX of Ref.~\cite{KSUS-19}. 
Nontrivial topological terms should influence the universal spectral statistics of non-Hermitian random matrices.
For example, in the Altland-Zirnbauer$^{\dag}$ class in Tables~\ref{tab: complex AZ} and \ref{tab: real AZ-dag}, three symmetry classes (i.e., classes AIII, BDI$^{\dag}$, and CII$^{\dag}$) exhibit $\mathbb{Z}$ topological terms while two symmetry classes (i.e., classes D$^{\dag}$ and DIII$^{\dag}$) accompany $\mathbb{Z}_2$ topological terms.
These five symmetry classes also coincide with the symmetry classes in which non-Hermitian random matrices support a subextensive number of real eigenvalues~\cite{Xiao-22}.
In this respect, it is also notable that a recent work~\cite{Garcia2023topo} investigates anomalous modes with real eigenvalues in dissipative quantum systems and introduces a relevant topological index.
Such anomalous real modes may be captured by our nonlinear sigma model description in the presence of topological terms.

%%%%% Hatano-Nelson %%%%%
\section{Hatano-Nelson model}
    \label{sec: Hatano-Nelson}

\begin{figure}[t]
\centering
\includegraphics[width=1.0\linewidth]{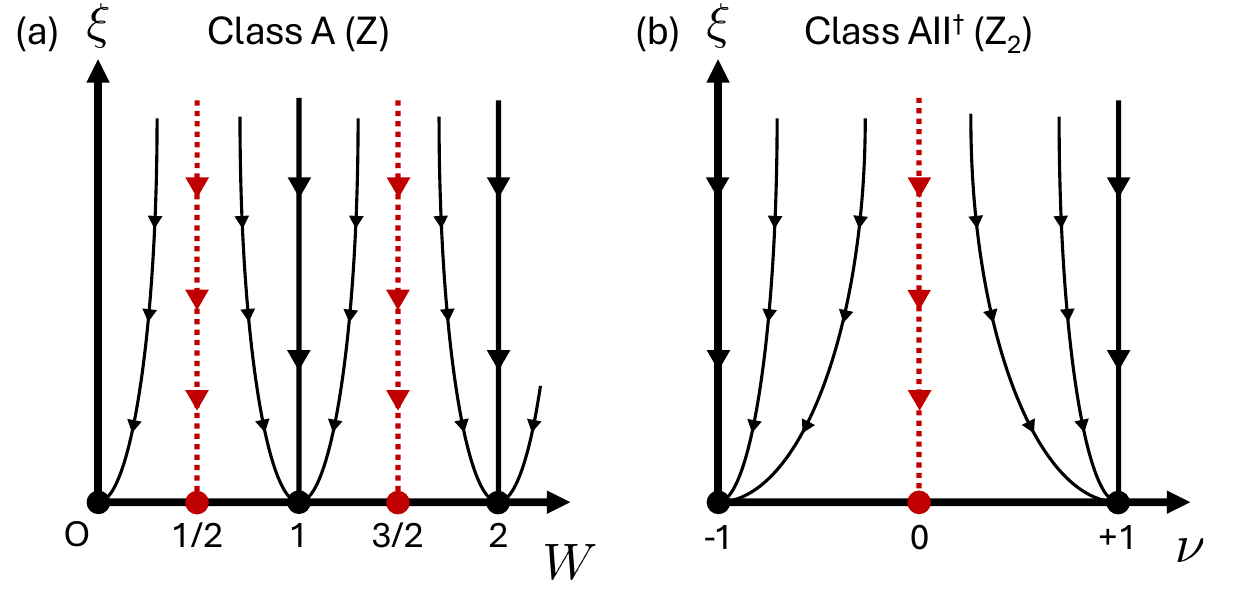} 
\caption{Renormalization-group flow of non-Hermitian disordered systems in one dimension.
(a)~Class A.
Two-parameter scaling based on the normal coupling parameter $\xi$ and the $\mathbb{Z}$ topological term $W$.
In addition to the fixed points with $\xi = 0$, $W \in \mathbb{Z}$ (black dots), the fixed points with $\xi = 0$, $W \in \mathbb{Z} + 1/2$ (red dots) appear.
(b)~Class AII$^{\dag}$.
Two-parameter scaling based on the normal coupling parameter $\xi$ and the $\mathbb{Z}_2$ topological term $\nu$.
In addition to the fixed points with $\xi = 0$, $\nu = \pm 1$ (black dots), the fixed point with $\xi = 0$, $\nu = 1/2$ (red dot) appears.}	
    \label{fig: 1D}
\end{figure}

In one dimension, Hermitian systems without symmetry are subject to Anderson localization even in the presence of infinitesimal disorder and hence exhibit no Anderson transitions~\cite{Abrahams-79, *Anderson-80}.
However, reciprocity-breaking non-Hermiticity can destroy Anderson localization and cause an Anderson transition even in one dimension~\cite{Hatano-Nelson-96, *Hatano-Nelson-97}. 
Here, we formulate a field-theoretical description of such a nonreciprocal Anderson transition.
In Sec.~\ref{subsec: HN-NLSM}, we derive the nonlinear sigma model,
\begin{equation}
    S_n [U] = \int dx \left[ \frac{\xi}{2} \mathrm{tr} \left[ \left( \partial_x U^{\dag} \right) \left( \partial_x U \right) \right] + W \mathrm{tr} \left[ U^{\dag} \partial_x U \right] \right],
        \label{eq: NLSM-1D}
\end{equation}
where the target manifold reads
\begin{equation}
    U \in \text{\Cone}.
\end{equation}
We specify the coupling constants $\xi$ and $W$ from the microscopic Dirac model.
In Sec.~\ref{subsec: HN-RG}, we study the renormalization-group flow of this nonlinear sigma model and show the Anderson transition due to the interplay of the kinetic and topological terms (Fig.~\ref{fig: 1D}).
We also clarify the physical meaning of the topological term and hence the anomalous delocalization in non-Hermitian systems.
Furthermore, in Sec.~\ref{subsec: HN-symplectic}, we discuss non-Hermitian disordered systems in the symplectic class (i.e., class AII$^{\dag}$) and develop a $\mathbb{Z}_2$ version of the nonlinear sigma model.

%%%%% NLSM %%%%%
\subsection{Derivation}
    \label{subsec: HN-NLSM}

The lattice Hamiltonian of the Hatano-Nelson model reads~\cite{Hatano-Nelson-96, *Hatano-Nelson-97},
\begin{align}
    &\hat{H} = \sum_{n} \bigg\{ - \frac{1}{2} \left[ \left( J+\gamma \right) \hat{\psi}_{n+1}^{\dag} \hat{\psi}_n + \left( J-\gamma \right) \hat{\psi}_{n}^{\dag} \hat{\psi}_{n+1} \right] \nonumber \\
    &\qquad\qquad\qquad\qquad\qquad\qquad\qquad+ m_n \hat{\psi}_n^{\dag} \hat{\psi}_n \bigg\},
\end{align}
where $J+\gamma \geq 0$ ($J-\gamma \geq 0$) denotes the hopping amplitude from the left to the right (from the right to the left), and $m_n \in \mathbb{C}$ is the random potential at site $n$.
To derive a nonlinear sigma model, it is useful to consider the continuum model,
$\hat{H} = \int dx~\hat{\psi}^{\dag} H \hat{\psi}$ with the non-Hermitian Dirac Hamiltonian~\cite{KR-21, KSR-21} 
\begin{equation}
    H = -\ii J \partial_x + \ii \gamma + V \left( x \right).
        \label{eq: HN-Dirac}
\end{equation}
This Dirac model describes one flavor of the lattice model at a Fermi point.
In both lattice and continuous descriptions, the degree $\gamma$ of non-Hermiticity plays a role of the imaginary-valued vector potential.
We here consider $N$ flavors of Dirac fermions and later take the large-$N$ limit, which enables the saddle-point analysis.
We assume the Gaussian random onsite potential $V \left( x \right) \in \mathbb{C}$, satisfying $\braket{V^{ij} \left( x \right)} = 0$ and $\braket{V^{ij} \left( x \right) \left( V^{kl} \left( x' \right) \right)^{*}} = \left( g^2/N \right) \delta^{ik} \delta^{jl} \delta \left( x-x' \right)$ with the disorder strength $g \geq 0$.
For simplicity, we set $J$ to unity in the following.

Following the procedure described in Sec.~\ref{subsec: replica}, we obtain the effective action $S_n \left[ Q \right]$, defined by $Z_n = \int \mathcal{D}Q~e^{-N S_n \left[ Q \right]}$, as
\begin{equation}
    S_n \left[ Q \right] = \int dx \left[ \frac{\mathrm{tr}\,Q^{\dag}Q}{g^2} - \mathrm{tr} \log \begin{pmatrix}
        G_0^{-1} & - Q \\
        Q^{\dag} & (G_0^{-1})^{\dag}
    \end{pmatrix} \right],
\end{equation}
where $Q = Q \left( x \right)$ is an $n \times n$ matrix field defined in one-dimensional space.
In the large-$N$ limit $N \to \infty$, the matrix integral is dominated by the saddle point of $S_n \left[ Q \right]$, 
\begin{equation}
    0 = \frac{\delta S_n}{\delta Q} \simeq \frac{Q^{\dag}}{g^2} - \left( QQ^{\dag} \right)^{-1} Q^{\dag}.
\end{equation}
As a result, $Q$ at the saddle point is generally represented as
\begin{equation}
    Q \left( x \right) = g U \left( x \right), \quad U \left( x \right) \in \mathrm{U} \left( n \right),
\end{equation}
which is consistent with our classification table (see class A in Table~\ref{tab: complex AZ}).
At the saddle point, the first term in $S_n \left[ Q \right]$ merely yields $n$, which we neglect in the following.

As shown in the classification table~\ref{tab: complex AZ}, we can have a topological term in the one-dimensional nonlinear sigma model for class A.
The Dirac model in Eq.~(\ref{eq: HN-Dirac}) realizes such a topological term giving rise to the unconventional Anderson transition.
To derive the topological term, we first change the basis of the Grassmann variables by $\psi' \coloneqq U^{\dag} \psi$, $\bar{\psi}' \coloneqq \bar{\psi} U$, which does not change the determinant.
Then, $S_n \left[ U \right]$ reduces to
\begin{align}   
    S_n \left[ U \right] &= -\int dx~\mathrm{tr} \log \begin{pmatrix}
        G_0^{-1} + \ii U^{\dag} \partial_x U & -g \\
        g & (G_0^{-1})^{\dag}
    \end{pmatrix} \nonumber \\
    &= -\int dx~\mathrm{tr} \log \left[ \mathcal{G}_0^{-1} + \begin{pmatrix}
        \ii U^{\dag} \partial_x U & 0 \\
        0 & 0
    \end{pmatrix} \right]
\end{align}
with 
\begin{align}
    \mathcal{G}_0^{-1} &\coloneqq \begin{pmatrix}
        G_0^{-1} & -g \\
        g & (G_0^{-1})^{\dag}
    \end{pmatrix} \nonumber \\
    &= \begin{pmatrix}
        E + \ii \partial_x - \ii \gamma & -g \\
        g & E^{*} + \ii \partial_x + \ii \gamma
    \end{pmatrix}.
\end{align}
Now, we assume that the spatial variation of $U \left( x \right)$ is sufficiently small and expand the action $S_n \left[ U \right]$ with respect to $U^{\dag} \partial_x U$, leading to
\begin{align}
    &S_n \left[ U \right] \simeq - \int dx~\mathrm{tr} \left[ \mathcal{G}_0 \begin{pmatrix}
        \ii U^{\dag} \partial_x U & 0 \\
        0 & 0
    \end{pmatrix} \right. \nonumber \\ 
    &\quad \left. - \frac{1}{2} \mathcal{G}_0 \begin{pmatrix}
        \ii U^{\dag} \partial_x U & 0 \\
        0 & 0
    \end{pmatrix} \mathcal{G}_0 \begin{pmatrix}
        \ii U^{\dag} \partial_x U & 0 \\
        0 & 0
    \end{pmatrix} \right] + \mathrm{const.}
\end{align}
In this Dirac model, the real part of the single-particle energy $E \in \mathbb{C}$ merely shifts momentum while the imaginary part the non-Hermitian term $\gamma$.
Hence, we below omit the $E$ dependence and set $E$ to zero.

The first-order term gives rise to the topological term in Eq.~(\ref{eq: NLSM-1D}), in a similar manner to the topological field theory description in Ref.~\cite{KSR-21}.
It reduces to 
\begin{equation}
    W \int dx~\mathrm{tr} \left[ U^{\dag} \partial_x U \right],
\end{equation}
where $W$ is evaluated as the momentum integral
\begin{align}
    W \coloneqq \int_{-\Lambda}^{\Lambda} \frac{dk}{2\pi\ii} \mathrm{tr} \left[ \mathcal{G}_0 \left( k \right) \begin{pmatrix}
        1 & 0 \\ 0 & 0
    \end{pmatrix} \right].
        \label{eq: 1D winding}
\end{align}
Here, $\Lambda > 0$ is a momentum cutoff that regularizes the Dirac model in Eq.~(\ref{eq: HN-Dirac}), and $\mathcal{G}_0 \left( k \right)$ is the Green's function in momentum space, given as
\begin{align}
    \mathcal{G}_0^{-1} \left( k \right) &= \begin{pmatrix}
        - k - \ii \gamma & -g \\
        g & - k + \ii \gamma
    \end{pmatrix}, \\
    \mathcal{G}_0 \left( k \right) &= - \frac{1}{k^2 + \gamma^2 + g^2} \begin{pmatrix}
        k - \ii \gamma & -g \\
        g & k + \ii \gamma
    \end{pmatrix}.
\end{align}
Then, the momentum integral is explicitly performed as
\begin{align}
    W &= \frac{\mathrm{sgn}\,\gamma}{\pi \sqrt{1 + \left( g/\gamma \right)^2}} \tan^{-1} \left( \frac{\Lambda}{\gamma^2 + g^2} \right) \nonumber \\
    &\rightarrow \frac{\mathrm{sgn}\,\gamma}{2 \sqrt{1 + \left( g/\gamma \right)^2}} \quad \left( \Lambda \to \infty \right).
\end{align}
In the absence of disorder (i.e., $g=0$), $W$ is quantized to be $\left( \mathrm{sgn}\,\gamma\right)/2$ and reduces to the integer topological invariant in terms of a point gap with respect to $E=0$~\cite{Gong-18, KSUS-19}.
The disorder $g \neq 0$ smoothens the sharp behavior of the sign function.
Thus, Eq.~(\ref{eq: 1D winding}) gives a disordered generalization of the point-gap topological invariant.
While such a topological invariant is quantized to be an integer in an individual sample even in the presence of disorder, $W$ corresponds to its statistical average for many realizations of the sample, leading to the deviation from the quantization.
Similar formulas of the complex-spectral winding number in the presence of disorder were also provided in Refs.~\cite{Moustaj-22, Nakai-24}.

On the other hand, the second-order term leads to the kinetic term
\begin{equation}
    \frac{\xi}{2} \int dx~\mathrm{tr} \left[ \left( \partial_x U^{\dag} \right) \left( \partial_x U \right) \right],
\end{equation}
where the coupling constant $\xi$ reads
\begin{align}
    \xi &= \int_{-\Lambda}^{\Lambda} \frac{dk}{2\pi} \left( \left[ \mathcal{G}_0 \left( k \right) \right]_{1,1} \right)^2 \nonumber \\
    %&= \int_{-\Lambda}^{\Lambda} \frac{dk}{2\pi} \frac{k^2 - \gamma^2}{\left( k^2 + \gamma^2 + g^2 \right)^2} \nonumber \\
    &= \frac{g^2}{\left( \gamma^2 + g^2\right)^{3/2}} \tan^{-1} \left( \frac{\Lambda}{\sqrt{\gamma^2 + g^2}}\right) \nonumber \\
    &\qquad\qquad\qquad\quad - \frac{\left( 2\gamma^2 + g^2 \right) \Lambda}{\left( \gamma^2 + g^2 \right) \left( \gamma^2 + g^2 + \Lambda^2 \right)} \nonumber \\
    &\rightarrow \frac{\pi g^2}{2\left( \gamma^2 + g^2\right)^{3/2}} \quad \left( \Lambda \to \infty \right).
\end{align}
Here, $\xi$ gives a length scale in the nonlinear sigma model.

%%%%% RG %%%%%
\subsection{Renormalization-group flow}
    \label{subsec: HN-RG}

The nonlinear sigma model obtained in Eq.~(\ref{eq: NLSM-1D}) is identical to that for one-dimensional Hermitian systems with chiral symmetry (i.e., class AIII).
Consequently, the renormalization-group flow should also be common.
Specifically, the renormalization-group equations are formulated as~\cite{Altland-14, *Altland-15}
\begin{align}
    G &= \sqrt{\frac{2\xi}{\pi L}} \sum_{l \in \mathbb{Z} +1/2} e^{-(l-W)^{2}L/2\xi}, \\
    W &= n - \frac{1}{4} \sum_{l \in \mathbb{Z} + 1/2} \left[ \mathrm{erf} \left( \sqrt{\frac{L}{2\xi}} \left( l-\delta W \right)^2 \right) \right. \nonumber \\
    &\qquad\qquad\qquad\qquad \left. - \mathrm{erf} \left( \sqrt{\frac{L}{2\xi}} \left( l+\delta W \right)^2 \right) \right],
\end{align}
where $G$ represents the conductance of the corresponding Hermitized system, and $\delta W \coloneqq W - n$ denotes the deviation of $W$ from the nearest integer $n \in \mathbb{Z}$.
In Fig.~\ref{fig: 1D}\,(a), we present the resultant renormalization-group flow based on the two parameters $\xi$ and $W$.
This is also compatible with the two-parameter scaling derived from the scattering theory of a microscopic non-Hermitian Hamiltonian~\cite{KR-21}.
In the absence of the topological term (i.e., $W = 0$), only a single fixed point $\xi = 0$, $W=0$ for the localized phase is relevant, which is a manifestation of the absence of delocalization in one dimension based on the single-parameter scaling hypothesis~\cite{Abrahams-79, *Anderson-80}.
By contrast, the nontrivial topological term $W \neq 0$ induces additional fixed points, stable ones $\xi = 0$, $W \in \mathbb{Z}$ and unstable ones $\xi = 0$, $W \in \mathbb{Z} + 1/2$.
This enables the topological Anderson transitions even in one dimension between the different phases characterized by the topological terms $W_0 \in \mathbb{Z}$ at the stable fixed points.

A crucial distinction from the corresponding Hermitized case lies in the physical significance of the topological term $W$.
As also discussed above, $W$ quantifies the winding of the complex spectrum.
Specifically, for a given disordered non-Hermitian Hamiltonian $H \left( \phi \right)$ with a $\mathrm{U} \left( 1 \right)$ flux $\phi \in \left[ 0, 2\pi \right)$, the point-gap topological invariant $W_0 \left( E \right) \in \mathbb{Z}$ is defined as~\cite{Gong-18, KSUS-19}
\begin{equation}
    W_0 \left( E \right) = \oint_0^{2\pi} \frac{d\phi}{2\pi\ii} \frac{d}{d\phi} \log \det \left[ H \left( \phi \right) - E \right]
        \label{eq: 1D topo inv}
\end{equation}
for a single-particle reference energy $E \in \mathbb{C}$.
Notably, nonzero $W_0 \neq 0$ implies the delocalization of wave functions and thus the metal phase, while $W = 0$ the localization and insulator phase.
This is in a similar spirit to the Thouless criterion~\cite{Thouless-review}, which characterizes localization through the insensitivity of the spectrum to the boundary conditions.
Consequently, the Anderson transition between the fixed point with $W=0$ and that with $W = \pm 1$ occurs between the insulator and metal phases.
This sharply contrasts with the Hermitized case, in which the Anderson transition separates the two different insulator phases since the topological term $W$ represents the electric polarization~\cite{Vanderbilt-textbook}.
In passing, we note that the microscopic topological invariant $W_0$ in Eq.~(\ref{eq: 1D topo inv}) is quantized to be integers whereas the topological term $W$ in Eq.~(\ref{eq: NLSM-1D}) is not necessarily quantized;
$W$ is extracted by averaging $W_0$ over disordered ensembles for a given system size $L$.

We clarify the physical relevance of the topological term on the basis of the nonlinear sigma model in Eq.~(\ref{eq: NLSM-1D}), in a similar manner to Refs.~\cite{Nazarov-94, LSZ-04, Altland-05, Ryu-07-Landauer}.
Akin to the formulation of the topological invariant in Eq.~(\ref{eq: 1D topo inv}), we insert a $\mathrm{U} \left( 1 \right)$ flux $\phi \in \left[ 0, 2\pi \right)$ and assume a uniform gauge potential $A = \phi/L$.
The unitary matrix $U$ as a target manifold can be expressed as
\begin{equation}
    U = \mathrm{diag}\,( e^{\ii \phi x/L} ) \in \mathrm{U} \left( n \right).
\end{equation}
Then, Eq.~(\ref{eq: NLSM-1D}) is given as
\begin{equation}
    S_n = \left( \frac{\xi \phi^2}{2L} + \ii W\phi \right) n,
\end{equation}
and the partition function $\Phi$ in Eq.~(\ref{eq: Phi - ZnE}) is 
\begin{equation}
    \Phi = \lim_{n\to 0} \frac{e^{-S_n} - 1}{n} = - \left( \frac{\xi \phi^2}{2L} + \ii W\phi \right).
\end{equation}
Thus, the kinetic and topological terms are extracted from $\Phi$ by
\begin{equation}
    \xi = - L \frac{\partial^2 \Phi}{\partial \phi^2}, \qquad
    W = \ii \frac{\partial \Phi}{\partial \phi}.
\end{equation}
From the definition of $\Phi$ in Eq.~(\ref{eq: Phi}), we further obtain
\begin{align}
    W = \ii \Braket{\tr \left[ \left( E-H \right)^{-1} \frac{\partial H}{\partial \phi} \right]+ \mathrm{c.c.}},
\end{align}
which can be interpreted as a certain kind of current in non-Hermitian systems.
This is compatible with the topological field theory description in Ref.~\cite{KSR-21}.
Moreover, it merits further study to explore its connection with the imaginary current introduced in Ref.~\cite{Hatano-Nelson-96, *Hatano-Nelson-97}.

%%%%% skin %%%%%
\subsection{Non-Hermitian skin effect}

In the presence of boundaries, the topological term of the nonlinear sigma model in Eq.~(\ref{eq: NLSM-1D}) is no longer invariant under the gauge transformation.
In fact, it reduces to 
\begin{equation}
    W \int_0^L dx~\mathrm{tr} \left[ U^{\dag} \partial_x U \right] = W \log \frac{\det U \left( L \right)}{\det U \left( 0 \right)}.
\end{equation}
The multivaluedness of the logarithm leads to the explicit dependence on the gauge choice.
This gauge dependence necessities the emergence of the boundary degree of freedom that restores the gauge invariance (i.e., anomaly inflow~\cite{KSR-21}).
Notably, such boundary modes appear for arbitrary single-particle energy $E$ with $W \neq 0$, and hence an extensive number of eigenmodes are localized.
This is a nonlinear sigma model description of the non-Hermitian skin effect~\cite{Lee-16, YW-18-SSH, Kunst-18, Yokomizo-19, Zhang-20, OKSS-20}.
The localization of an extensive number of eigenmodes contrasts with the corresponding Hermitian systems with chiral symmetry, in which only zero modes host the topological term and are subject to localization.

%%%%% 1D class AII^{\dag} %%%%%
\subsection{Symplectic Hatano-Nelson model (class AII$^{\dag}$)}
    \label{subsec: HN-symplectic}

A symplectic extension of the Hatano-Nelson model is formulated as~\cite{OKSS-20, KR-21}
\begin{align}
    &\hat{H} = \sum_{n} \bigg\{ - \frac{1}{2} \left[ \hat{\psi}_{n+1}^{\dag} \left( J + \gamma \sigma_z - \ii \Delta \sigma_x \right) \hat{\psi}_n  \right. \nonumber \\
    &\qquad \left. + \left( J - \gamma \sigma_z + \ii \Delta \sigma_x \right) \hat{\psi}_{n}^{\dag} \hat{\psi}_{n+1} \right]+ m_n \hat{\psi}_n^{\dag} \hat{\psi}_n \bigg\},
\end{align}
where the fermionic annihilation operator $\hat{\psi}_n = (\hat{\psi}_{n, \uparrow}, \hat{\psi}_{n, \downarrow})^T$ [creation operator $\hat{\psi}_n^{\dag} = (\hat{\psi}_{n, \uparrow}^{\dag}, \hat{\psi}_{n, \downarrow}^{\dag})$] now encompasses the spin degree of freedom, and $\sigma_i$'s ($i=x, y, z$) represent Pauli matrices.
Here, non-Hermiticity $\gamma > 0$ ($\gamma < 0$) drives the up-spin fermions $\hat{\psi}_{n, \uparrow}$'s toward the right (left) but the down-spin fermions $\hat{\psi}_{n, \downarrow}$'s toward the left (right).
Additionally, $\Delta \in \mathbb{R}$ describes the spin-orbit coupling between the up-spin and down-spin fermions.

Unlike the original Hatano-Nelson model, nonreciprocity is preserved on average, although each of up-spin or down-spin fermions individually breaks nonreciprocity.
Consequently, the symplectic Hatano-Nelson model respects time-reversal symmetry$^{\dag}$ in Eq.~(\ref{eq: TRSdag}), where $\mathcal{T}$ is given as $\mathcal{T} = \sigma_y$, and thus belongs to class AII$^{\dag}$ within the 38-fold symmetry classification~\cite{KSUS-19}.
It exhibits the Anderson transition arising from the interplay of non-Hermiticity, spin-orbit coupling, and disorder~\cite{KR-21}.

As shown in Table~\ref{tab: real AZ-dag}, the symplectic Hatano-Nelson model is effectively described by the one-dimensional nonlinear sigma model with the target manifold $\mathrm{O} \left( n \right)$.
Notably, akin to the corresponding Hermitized systems in class DIII~\cite{Altland-14, *Altland-15}, this model features a $\mathbb{Z}_2$ topological term $\nu$, which underlies the Anderson transition.
We accordingly present the consequent renormalization-group flow in Fig.~\ref{fig: 1D}\,(b), consistent with the two-parameter scaling derived from a microscopic model~\cite{KR-21}.
Other symmetry classes that host the $\mathbb{Z}_2$ topological term are identified in Tables~\ref{tab: complex AZ}-\ref{tab: real AZ + SLS}.

%%%%% others %%%%%
\section{Universality classes in two and three dimensions}
    \label{sec: others}

We discuss the Anderson transitions for other symmetry classes in two and three dimensions, relying on the nonlinear sigma model description.
Here, we mainly focus on the consequences of the potential topological terms for exemplary universality classes.

%%%%% 2D class AIII %%%%%
\subsection{Quantum Hall transition (2D class AIII; 2D classes CII and CI$^{\dag}$)}

While two-dimensional Hermitian systems without symmetry (i.e., class A) 
are subject to Anderson localization even in the presence of infinitesimal disorder~\cite{Abrahams-79}, nontrivial topology gives rise to the unique Anderson transitions.
Prime examples include the quantum Hall transition~\cite{QHE-textbook, Huckestein-review, Kramer-review, Evers-review}.
From the field-theoretical perspective, the quantum Hall transition is formulated by the two-parameter scaling of the nonlinear sigma model with the topological term (i.e., $\theta$ term)~\cite{Khmel'nitskii-83, Levine-83, *Pruisken-84, *Pruisken-88}.
Recent development is found, for example, in Refs.~\cite{Slevin-09, Obuse-12, Zirnbauer-19, Sbierski-20, Dresselhaus-21, Dresselhaus-22}.

Owing to the correspondence between Hermitian and non-Hermitian systems, non-Hermitian counterparts of the quantum Hall transition occur in certain symmetry classes, which are identified by our classification tables~\ref{tab: complex AZ}-\ref{tab: real AZ + SLS}.
As an exemplary class, we here focus on two-dimensional non-Hermitian systems in class AIII. As a defining feature, they are required to respect chiral symmetry
\begin{equation}
    \Gamma H^{\dag} \Gamma^{-1} = - H
\end{equation}
with a unitary matrix $\Gamma$ satisfying $\Gamma^2 = 1$.
As a result of chiral symmetry, the replica partition function $Z_n$ in Eq.~(\ref{eq: replica partition function}) reduces to that of Hermitized systems in class A.
Specifically, the effective nonlinear sigma model is given as
\begin{equation}
    S_n [Q] = \frac{1}{t} \sum_{i=x, y} \int d^2x~\mathrm{tr} \left[ (\partial_i Q^{\dag}) (\partial_i Q) \right] + \theta N [Q]
\end{equation}
with the topological term
\begin{equation}
    N [Q] \coloneqq \sum_{i, j = x, y} \varepsilon^{ij} \int \frac{d^2 x}{16\pi} \mathrm{tr} \left[ Q (\partial_i Q) (\partial_j Q)\right],
\end{equation}
where the saddle-point target manifold reads
\begin{equation}
    Q \in \text{\Czero}.
\end{equation}
It should be noted that this nonlinear sigma model is valid only for complex energy satisfying chiral symmetry (i.e., pure imaginary energy $E \in \ii \mathbb{R}$), whereas it is applicable to arbitrary real-valued energy in the Hermitian case.
For microscopic non-Hermitian disordered models, the integer-valued point-gap topological invariant is given as the Chern number of the Hermitian matrix $\ii H\Gamma$~\cite{KSUS-19}.

\begin{figure}[t]
\centering
\includegraphics[width=0.6\linewidth]{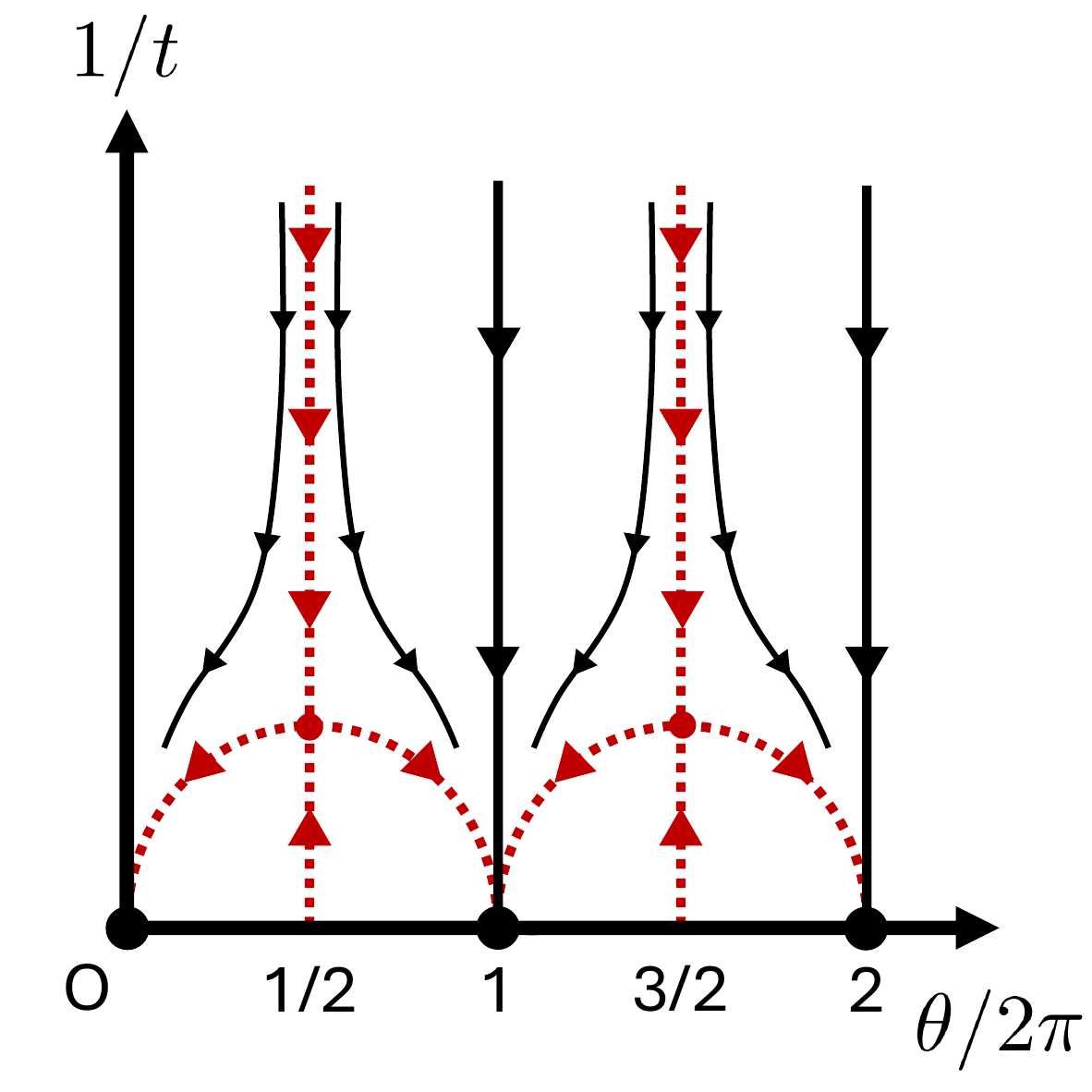} 
\caption{Renormalization-group flow of non-Hermitian disordered systems in two dimensions for class AIII based on the normal coupling parameter $t$ and the topological theta term $\theta$.
In addition to the fixed points with $1/t = 0$, $\theta/2\pi \in \mathbb{Z}$ (black dots), the fixed points with $1/t > 0$, $\theta/2\pi \in \mathbb{Z} + 1/2$ (red dots) appear.}	
    \label{fig: 2D-A}
\end{figure}

Because of the common underlying nonlinear sigma model, two-dimensional non-Hermitian disordered systems in class AIII are also described by the two-parameter scaling (Fig.~\ref{fig: 2D-A}).
In addition to the trivial fixed points with $1/t=0$, $\theta/2\pi \in \mathbb{Z}$, the nontrivial fixed points with $1/t > 0$, $\theta/2\pi \in \mathbb{Z} + 1/2$ appear, which should accompany Anderson transitions.
In the original quantum Hall transition for Hermitian systems, $1/t$ and $\theta$ physically correspond to the longitudinal and Hall conductivity, respectively.
However, non-Hermiticity can change their physical meanings, as is also the case for one-dimensional systems (see Sec.~\ref{sec: Hatano-Nelson}).
The two-parameter scaling developed here should underlie the magnon Hall transition in Ref.~\cite{Xu-16}.
Other possible lattice models with nontrivial point-gap topology (but without disorder) were also provided, for example, in Ref.~\cite{Nakamura-24}.

Furthermore, it is notable that the quantum Hall transition is effectively described by network models, including the Chalker-Coddington model~\cite{CC-88}.
These models were also generalized to superconducting counterparts of the quantum Hall transitions in classes C and D, which respectively correspond to classes CII and CI$^{\dag}$ and classes BDI and DIII$^{\dag}$ in the non-Hermitian regime (see Tables~\ref{tab: real AZ} and \ref{tab: real AZ-dag}).
Important advantages of such network models include analytical tractability.
For example, the spin quantum Hall transition in disordered superconductors in class C can be mapped to the two-dimensional percolation, enabling us to obtain the exact values of the critical exponents~\cite{Kagalovsky-99, Gruzberg-99}.
It merits further research to construct non-Hermitian counterparts of the network models and investigate them with various approaches.

%%%%% 2D class A %%%%%
\subsection{Gade-Wegner nonlinear sigma model (2D class A)}

In two-dimensional Hermitian disordered systems with chiral symmetry (i.e., class AIII), the beta function for conductivity vanishes strictly, and the corresponding coupling parameter is not renormalized, further implying the absence of the Anderson localization.
This follows from the two-dimensional nonlinear sigma model with the target manifold \Cone~with the additional Gade-Wegner term~\cite{Gade-91, *Gade-93}.
According to the correspondence between Hermitian and non-Hermitian systems, such a Gade-Wegner nonlinear sigma model has a non-Hermitian counterpart in two-dimensional systems without symmetry (i.e., class A).

%%%%% 3D Wigner-Dyson %%%%%
\subsection{3D Wigner-Dyson class (classes A, AI$^{\dag}$, and AII$^{\dag}$)}

In three dimensions, non-Hermitian systems in the Wigner-Dyson class (i.e., classes A, AI$^{\dag}$, and AII$^{\dag}$) are characterized by the integer-valued point-gap topological invariant (i.e., three-dimensional complex-spectral winding number)~\cite{KSUS-19}.
The corresponding nonlinear sigma models reduce to those for three-dimensional Hermitian systems in classes AIII, CI, and DIII, reading~\cite{Ryu-12}
\begin{equation}
    S_n [Q] = \frac{1}{t} \sum_{i=x, y, z} \int d^3x~\mathrm{tr} \left[ (\partial_i Q^{\dag}) (\partial_i Q) \right] + \ii \theta \Gamma [Q]
\end{equation}
with the topological term
\begin{align}
    &\Gamma [Q] \coloneqq \sum_{i, j, k = x, y, z} \varepsilon^{ijk} \int \frac{d^3 x}{24\pi^2} \nonumber \\
    &\qquad\qquad\qquad \times \mathrm{tr} \left[ Q^{\dag} (\partial_i Q) Q^{\dag} (\partial_j Q) Q^{\dag} (\partial_k Q)\right],
\end{align}
where the saddle-point target manifold is
\begin{equation}
    Q \in \begin{cases}
        \text{\Cone} & (\text{class A}); \\
        \text{\Rseven} & (\text{class AI}^{\dag}); \\
        \text{\Rthree} & (\text{class AII}^{\dag}).
    \end{cases}
\end{equation}

In comparison with lower dimensions, it is challenging to analyze the three-dimensional nonlinear sigma model in the presence of the topological term.
Possible normalization-group flows for three-dimensional Hermitian disordered systems in classes CI and DIII, corresponding to non-Hermitian systems in classes AI$^{\dag}$ and AII$^{\dag}$, were provided in Fig.~5 of Ref.~\cite{Ryu-12}.
Clean lattice models with the corresponding point-gap topology were provided in Refs.~\cite{Bessho-20, Denner-21, KSR-21, Nakamura-24}.
Additionally, while $\ii \theta \Gamma [Q]$ corresponds to strong point-gap topology, weak point-gap topology is equally relevant to the universality classes of Anderson transitions~\cite{Xiao-23, *Zhao-24}.

%%%%% Discussions %%%%%
\section{Discussions}
    \label{sec: discussions}

In this work, we formulate an effective field theory description of non-Hermitian disordered systems based on the fermionic replica nonlinear sigma models.
We comprehensively classify target manifolds of the nonlinear sigma models across all the 38-fold symmetry classes of non-Hermitian systems and elucidate the unique universality classes of the Anderson transitions in non-Hermitian disordered systems (Tables~\ref{tab: complex AZ}-\ref{tab: real AZ + SLS}).
Specifically, we develop a field-theoretical understanding of the duality between non-Hermitian systems and corresponding Hermitized systems with additional chiral symmetry, which was formulated for microscopic lattice models~\cite{Luo-22R}.
As an illustrative example of our field-theoretical framework, we study the Anderson transition of nonreciprocal disordered systems in one dimension, including the Hatano-Nelson model~\cite{Hatano-Nelson-96, *Hatano-Nelson-97}, and show that it arises from the competition between the kinetic and topological terms in the one-dimensional nonlinear sigma model.
We also discuss the potential consequences of topological terms and critical phenomena of non-Hermitian disordered systems for different symmetry classes in higher dimensions.

In addition to the replica approach, the supersymmetry~\cite{Efetov-textbook} and Keldysh~\cite{Kamenev-textbook} approaches are equally powerful in formulating effective field theory for Hermitian disordered systems.
For example, these approaches may yield the correct density of states for non-Hermitian random matrices, in contrast to the incorrect one in Eq.~(\ref{eq: DOS-D-incorrect}) obtained through the replica formalism.
It merits further research to explore such other approaches to investigate non-Hermitian disordered systems. 
Additionally, while we have focused on the one-point correlation function and its related quantities, including the density of states and the localization length, higher-order correlation functions are relevant to other physical quantities. 
As a prime example, the two-point correlation function characterizes electrical conductivity in Hermitian systems.
We leave it for future work to study higher-order correlation functions and related transport properties of non-Hermitian disordered systems from the field-theoretical perspective.
While we show that the Hermitization technique is pivotal for characterizing the one-point correlation function, its effectiveness in capturing higher-order correlation functions remains unclear.
Furthermore, nonlinear sigma model descriptions have recently been developed for the dynamics of free fermions subject to quantum measurements~\cite{Jian-22, *Jian-23, Yang-23, Fava-23, Poboiko-23}.
It is worthwhile to explore potential applications of our theory to such open quantum dynamics.

\bigskip
%%%%% Note added %%%%%
{\it Note added}.---When we were finalizing the draft, we learned about a related work~\cite{Akemann-24}.
After the initial submission of this work, we also learned about another related work~\cite{Forrester-24}.

%%%%% Acknowledgement %%%%%
\begin{acknowledgments}
We thank Zhenyu Xiao for helpful discussion.
K.K. thanks Xunlong Luo, Tomi Ohtsuki, Ryuichi Shindou, and Zhenyu Xiao for related collaborations.
K.K. is supported by MEXT KAKENHI Grant-in-Aid for Transformative Research Areas A ``Extreme Universe" No.~24H00945.
%S.R. is supported by Simons Investigator Grant from the  Simons Foundation (Grant No.~566116). 
This material is based upon work supported by the National Science Foundation under Grant No.\ DMR-2409412.
This work is supported by the Gordon and Betty Moore Foundation through Grant No.~GBMF8685 toward the Princeton theory program.
K.K., A.K., and S.R. gratefully acknowledge support from the Simons Center for Geometry and Physics, Stony Brook University at which some of the research for this work was performed.
\end{acknowledgments}

\appendix

%%%%%%%%%%
\section{Symmetry classification of Hermitian and non-Hermitian systems}
    \label{asec: symmetry}

We describe symmetry classification of Hermitian and non-Hermitian systems.
In Appendix~\ref{asec: symmetry - Hermitian}, we summarize the 10-fold classification of Hermitian systems based on time reversal~\cite{Wigner-51, *Wigner-58, Dyson-62}, charge conjugation (particle-hole transformation)~\cite{AZ-97}, and chiral transformation~\cite{Gade-91, *Gade-93, Verbaarschot-94, Verbaarschot-00-review}.
In Appendix~\ref{asec: symmetry - non-Hermitian}, we summarize the 38-fold classification of non-Hermitian systems~\cite{Bernard-LeClair-02, KSUS-19}.

%%%%%%%%%%
\subsection{10-fold symmetry classification of Hermitian systems}
    \label{asec: symmetry - Hermitian}

In general, Hermitian systems $H$ are classified according to the two types of antiunitary symmetry, one of which is time-reversal symmetry (TRS)
\begin{equation}
    \mathcal{T} H \mathcal{T}^{-1} = H,
        \label{aeq: TRS}
\end{equation}
and the other of which is particle-hole symmetry (PHS)
\begin{equation}
    \mathcal{C} H \mathcal{C}^{-1} = - H.
        \label{aeq: PHS-dag}
\end{equation}
Here, $\mathcal{T}$ and $\mathcal{C}$ are antiunitary operators satisfying $\mathcal{T}^2 =\pm 1$ and  $\mathcal{C}^2 = \pm 1$.
As a combination of TRS and PHS, we also introduce chiral symmetry (CS), or equivalently sublattice symmetry (SLS), by
\begin{equation}
    \mathcal{S} H \mathcal{S}^{-1} = - H,
        \label{aeq: SLS}
\end{equation}
where $\mathcal{S}$ is a unitary operator satisfying $\mathcal{S}^2 = 1$.
These two antiunitary symmetry and one unitary symmetry form the tenfold Altland-Zirnbauer (AZ) symmetry classification (Table~\ref{tab: Hermitian})~\cite{AZ-97}, determining the universality classes of Anderson transitions~\cite{Evers-review} and topological phases~\cite{CTSR-review} in Hermitian systems.
We note that $H$ is assumed not to have any unitary symmetry that commutes with it (i.e., $\mathcal{U} H \mathcal{U}^{-1} = H$); 
if $H$ of interest respects such unitary symmetry, we perform the block diagonalization and then study the internal symmetry in each subspace.

%%%%%%%%%%
\subsection{38-fold symmetry classification of non-Hermitian systems}
    \label{asec: symmetry - non-Hermitian}

In contrast to the 10-fold AZ symmetry classification for Hermitian systems, non-Hermitian systems are generally classified according to the 38-fold internal symmetry~\cite{KSUS-19}.
First, the two types of antiunitary symmetry in Eqs.~(\ref{aeq: TRS}) and (\ref{aeq: PHS-dag}) remain symmetry even for non-Hermitian systems $H$, each of which is respectively denoted by TRS and PHS$^{\dag}$~\cite{KSUS-19}.
Similarly, the unitary symmetry in Eq.~(\ref{aeq: SLS}) is still symmetry and denoted by SLS.
In addition to these symmetries, we consider symmetry that relates non-Hermitian systems $H$ to their Hermitian conjugates $H^{\dag}$.
We introduce two such antiunitary symmetries by
\begin{equation}
    \mathcal{T} H^{\dag} \mathcal{T}^{-1} = H,
        \label{aeq: TRS-dag}
\end{equation}
and
\begin{equation}
    \mathcal{C} H^{\dag} \mathcal{C}^{-1} = - H,
        \label{aeq: PHS}
\end{equation}
where $\mathcal{T}$ and $\mathcal{C}$ are antiunitary operators satisfying $\mathcal{T}^2 = \pm 1$ and $\mathcal{C}^2 = \pm 1$.
These symmetries are respectively denoted by time-reversal symmetry$^{\dag}$ (TRS$^{\dag}$) and particle-hole symmetry (PHS) because they are obtained by additional Hermitian conjugation to TRS and PHS$^{\dag}$ in Eqs.~(\ref{aeq: TRS}) and (\ref{aeq: PHS-dag}).
In a similar manner, we also consider Eq.~(\ref{aeq: SLS}) with additional Hermitian conjugation by
\begin{equation}
    \Gamma H^{\dag} \Gamma^{-1} = - H,
        \label{aeq: CS}
\end{equation}
where $\Gamma$ is a unitary operator satisfying $\Gamma^2 = 1$.
Unitary symmetry in Eq.~(\ref{aeq: SLS}) is called SLS for non-Hermitian systems while unitary symmetry in Eq.~(\ref{aeq: CS}) is called CS~\cite{KSUS-19}.
Here, while CS and SLS are equivalent to each other for Hermitian systems, they are different for non-Hermitian systems.
The combination of TRS and PHS, as well as the combination of TRS$^{\dag}$ and PHS$^{\dag}$, gives rise to CS;
on the other hand, the combination of TRS and PHS$^{\dag}$, as well as the combination of TRS$^{\dag}$ and PHS, gives rise to SLS.

Similarly to the 10-fold AZ symmetry class for Hermitian systems, TRS in Eq.~(\ref{aeq: TRS}), PHS in Eq.~(\ref{aeq: PHS}), and CS in Eq.~(\ref{aeq: CS}) form the 10-fold symmetry class for non-Hermitian systems (Tables~\ref{tab: complex AZ} and \ref{tab: real AZ}).
Moreover, TRS$^{\dag}$ in Eq.~(\ref{aeq: TRS-dag}), PHS$^{\dag}$ in Eq.~(\ref{aeq: PHS-dag}), and CS in Eq.~(\ref{aeq: CS}) form another 10-fold symmetry class, which is called the AZ$^{\dag}$ symmetry class for non-Hermitian systems (Tables~\ref{tab: complex AZ} and \ref{tab: real AZ-dag}).
In these AZ and AZ$^{\dag}$ symmetry classes, SLS in Eq.~(\ref{aeq: SLS}) is not included.
Taking SLS into consideration as additional symmetry (Tables~\ref{tab: complex AZ} and \ref{tab: real AZ + SLS}), we have the 38-fold symmetry class for non-Hermitian systems~\cite{KSUS-19}.

Some symmetry classes in Tables~\ref{tab: complex AZ}-\ref{tab: real AZ + SLS} give the equivalent symmetry classes, which are not double counted in the 38-fold symmetry classification.
For example, if a non-Hermitian system $H$ respects TRS in Eq.~(\ref{aeq: TRS}), another non-Hermitian system $\ii H$ respects PHS$^{\dag}$ in Eq.~(\ref{aeq: PHS-dag}), both of which exhibit essentially the same universal spectral statistics.
Consequently, classes AI and AII are respectively equivalent to classes D$^{\dag}$ and C$^{\dag}$ and characterized by the same universality classes (Table~\ref{tab: real AZ}).
See also Tables~XIII and XIV in Ref.~\cite{KSUS-19} for the correspondence between symmetry classes.

In non-Hermitian random matrix theory, TRS$^{\dag}$ in Eq.~(\ref{aeq: TRS-dag}) changes the spectral statistics in the bulk~\cite{Hamazaki-20}. 
On the other hand, TRS in Eq.~(\ref{aeq: TRS}), PHS$^{\dag}$ in Eq.~(\ref{aeq: TRS-dag}), and CS in Eq.~(\ref{aeq: CS}) are relevant to the spectral statistics around the symmetric lines (i.e., real and imaginary axes)~\cite{Xiao-22}, and PHS in Eq.~(\ref{aeq: PHS}) and SLS in Eq.~(\ref{aeq: SLS}) are relevant to those around the symmetric point (i.e., spectral origin)~\cite{GarciaGarcia-22PRX, Xiao-24}.
The 38-fold classification of non-Hermitian systems was also applied to the symmetry classification of quadratic~\cite{Lieu-20} and many-body~\cite{Sa-23, Kawabata-23} Lindbladians.

\begin{table*}[t]
    \centering
    \footnotesize
    \caption{Characteristic polynomials of non-Hermitian random matrices.
    $Z_n \left( z \right)$ [$Z^{\mathrm{b}}_{n}\left( z \right)$] is the one-point fermionic (bosonic) replica partition function, where $z \in \mathbb{C}$ corresponds to $E \in \mathbb{C}$ in the text.
    When the complex spectrum of the ensemble is rotationally invariant around the origin (i.e., classes A, $\mathrm{AI}^\dagger$, $\mathrm{AII}^\dagger$, D, and C), $Z_{n}(z)$ and $Z^{\mathrm{b}}_{n}(z)$ depend only on $\left| z \right|$.
    When the complex spectrum exhibits a symmetric line (i.e., classes AI and AII), $Z_{n}(z)$ is a product of factors dependent on $x \coloneqq \operatorname{Re}\,z$ and $y \coloneqq \operatorname{Im}\,z$.
    $Z_n \left( z_1, z_2 \right)$ is the two-point fermionic replica partition function as a function of $z \coloneqq \left( z_1+z_2 \right)/2$ and $\omega \coloneqq  z_1-z_2$.}
    \begin{tabular}{cccccccc}
    \hline \hline
    Class & Definition & \multicolumn{2}{c}{Parameters} & Manifold & Symmetry & \multicolumn{1}{c}{Characteristic polynomial} & Ref. \\
    \hline
    A & $H\in \mathbb{C}^{N\times N}$ & 1pt & $\forall\,N$ & $\mathbb{C}^{n\times n}$ & $\mathrm{U} \left( n \right)$ & $Z_n \left(z \right) = \left| z \right|^{2n(N+n)} \mathcal{J}^{(1,N+1,1)}_n \,( |z|^2 )$ & \cite{Nishigaki-02} \\
    & & 2pt & $N\to\infty$ & $\mathrm{U}\left( 2n \right)$ & $\mathrm{U} \left( n \right)\times \mathrm{U} \left( n \right)$ & $Z_n \left( z_1, z_2 \right) = e^{2n \left| z \right|^2} e^{-n|\omega|^2/2} I^{(1,1,1)}_n \,( |\omega|^2 )$ & \cite{Nishigaki-02} \\ \hline
    $\mathrm{AI}^\dagger$ & $H^T = H$ & 1pt & $\forall\,N$ & $\mathbb{H}^{n\times n}$ & $\mathrm{Sp} \left(n \right)$ & $Z_n \left(z \right) = \left| z \right|^{2n(N+2n)}\mathcal{J}^{(2,N+1,2)}_n \,(2|z|^2 )$ & \cite{Anish} \\
    & & 2pt & $N\to\infty$ & $\mathrm{Sp}\left( 2n \right)$ & $\mathrm{Sp} \left( n \right)\times \mathrm{Sp} \left( n \right)$ & $Z_n \left( z_1, z_2 \right) = e^{4n|z|^2}e^{n|\omega|^2} I^{(2,2,2)}_n \,( -2|\omega|^2 )$ & \cite{Anish} \\ \hline
    $\mathrm{AII}^\dagger$ & $\sigma_y H^T \sigma_y = H$ & 1pt & $\forall\,N$ & $\mathbb{R}^{2n\times 2n}$ & $\mathrm{O}\left( 2n \right)$ & $Z_n \left( z \right) = \left| z \right|^{2n(N+2n)} \mathcal{J}^{(1/2,N+1,1/2)}_{2n}\,( |z|^2 )$ & \cite{Anish} \\
    & & 2pt & $N\to\infty$ & $\mathrm{O}\left( 4n \right)$ & $\mathrm{O} \left( 2n \right) \times \mathrm{O} \left( 2n \right)$ & $Z_n \left( z_1, z_2 \right) = e^{4n|z|^2} e^{n|\omega|^2} I^{(1/2,1/2,1/2)}_{2n} \,( -|\omega|^2 )$ & \cite{Anish} \\ \hline
    AI & $H^* = H$ & 1pt & $N\to\infty$, even $n$ & \multirow{2}{*}{$\mathrm{U}\left( 2n \right)/\mathrm{Sp} \left( n \right)$} & \multirow{2}{*}{$\mathrm{U} \left( n \right)\times \mathrm{U} \left( n \right)$} & $Z_n \left( z \right) = e^{n|z|^2} I^{(1,2,2)}_{n/2}\,( -4y^2 )$ \\
    & & 1pt & $N\to\infty$, odd $n$ & & & $Z_n \left( z \right) = e^{n|z|^2} I^{(3,2,2)}_{(n-1)/2}\,( -4y^2 )$ \\
    & & 1pt & $N\to\infty$, bosonic & $\mathrm{GL} \left( 2n,\mathbb{R} \right)/\mathrm{O} \left( 2n \right)$ & $\mathrm{O} \left( n \right) \times \mathrm{O} \left( n \right)$ & $Z^{\mathrm{b}}_{n} \left( z \right) = e^{-n|z|^2} \mathcal{J}^{(1,1/2,1/2)}_{n}\,( 2y^2 )$ & \cite{Nishigaki-02} \\ \hline
    AII & $\sigma_y H^* \sigma_y = H$ & 1pt & $N\to\infty$ & $\mathrm{U}\left( 2n \right)/\mathrm{O} \left( 2n \right)$ & $\mathrm{U} \left( n \right)\times \mathrm{U} \left( n \right)$ & $Z_n \left( z \right) = e^{n|z|^2} I^{(1,1/2,1/2)}_n\,(-2y^2)$ & \cite{Nishigaki-02} \\ \hline
    D & $H^T = -H$ & 1pt & $N\to\infty$, even $n$ & \multirow{4}{*}{$\mathrm{O}\left( 2n \right)$} & \multirow{4}{*}{$\mathrm{U} \left( n \right)$} & $Z_n \left( z \right) = e^{n|z|^2}I^{(1,1,2)}_{n/2}\,(-4|z|^2)/2$ & Eq.~(\ref{eq: class d saddle pt even}) \\
    & & & & & & $\qquad + e^{(n-2)|z|^2}I^{(3,3,2)}_{n/2-1}\,(-4|z|^2) / 2$ \\
    & & 1pt & $N\to\infty$, odd $n$ & & & $Z_n \left( z \right) = e^{(n-2)|z|^2}I^{(1,3,2)}_{(n-1)/2}\,(-4|z|^2)/2$ & Eq.~(\ref{eq: class d saddle pt odd}) \\
    & & & & & & $\qquad + e^{n|z|^2}I^{(3,1,2)}_{(n-1)/2}\,(-4|z|^2)/2$ \\ \hline
    C & $\sigma_y H^T \sigma_y = -H$ & 1pt &  $N\to\infty$ & $\mathrm{Sp} \left( n \right)$ & $\mathrm{U} \left( n \right)$ & $Z_n \left( z \right) = e^{n|z|^2} I^{(1,1,1/2)}_n\,(-2|z|^2)$ 
    &
    Eq.\ \eqref{class c result}
    \\
    \hline \hline
\end{tabular}
    \label{tab: nh rmt known results}
\end{table*}

\begin{table*}[t]
    \centering
    \footnotesize
    \caption{Comparison of characteristic polynomials of non-Hermitian random matrices for small $N$ and at saddle points.
    For a fixed number $n$ of replicas, the characteristic polynomials $Z_n \left( z \right)$ for small matrix size $N$ are computed either analytically (i.e., by integral from the definition) or numerically (i.e., by taking the average of characteristic polynomials of samples).
    %In classes D and C, $z \in \mathbb{C}$ corresponds to $2E \in \mathbb{C}$ in the text.
    For $n=3,4$ and $N=4,6$ in class D, $Z_n \left( z \right)$ is numerically obtained from $10^9$ realizations of non-Hermitian random matrices.
    In class AI, we introduce $x \coloneqq \mathrm{Re}\,z$ and $y \coloneqq \mathrm{Im}\,z$ for $z \in \mathbb{C}$.
    The low-degree terms of $Z_n \left( z \right)$ for small $N$ agree with those of $Z_n \left( z \right)$ at the saddle points.
    As $N$ increases, the number of terms in agreement also increases.}
    \renewcommand{\arraystretch}{1.5} 
\begin{tabular}{cccll}
    \hline \hline
    Class & Replica $n$ & Size $N$ & $Z_n \left( z \right)$ of small $N$ & $Z_n \left( z \right)$ at saddle point \\
    \hline
    \multirow{16}{*}{D} & $1$ & even $N$ & $\sum_{k=0,2,\cdots,N} \frac{1}{k!} \left({ |z|^{2}}\right)^{k}$ & $\sum_{k=0,2,\cdots,\infty} \frac{1}{k!} \left({ |z|^{2}}\right)^{k}$ \\
    \cline{2-5}
     & \multirow{3}{*}{$2$} & $N=2$ & $1+\frac{ |z|^{4}}{3}+\frac{ |z|^{8}}{24}$ & \multirow{3}{*}{$1+\frac{ |z|^{4}}{3}+\frac{ |z|^{8}}{15} + \frac{2 |z|^{12}}{315} + \cdots$} \\
     & & $N=4$ & $1+\frac{ |z|^{4}}{3}+\frac{ |z|^{8}}{15}+\frac{ |z|^{12}}{180} + \cdots$ & \\
     & & $N=6$ & $1+\frac{ |z|^{4}}{3}+\frac{ |z|^{8}}{15} + \frac{2 |z|^{12}}{315} + \cdots$ \\
     \cline{2-5}
     & \multirow{4}{*}{$3$} & $N=2$ & $1+\frac{3 |z|^{4}}{10}+\frac{ |z|^{8}}{40}+\cdots$ & \multirow{4}{*}{$\begin{array}{l}1+\frac{3 |z|^{4}}{10}+\frac{13 |z|^{8}}{280} + \frac{79 |z|^{12}}{15120}+\cdots \\ \scriptstyle = 1 + \num{0.3}|z|^4 + \num{0.0464286}|z|^8 + \num{0.00522487}|z|^{12} + \cdots\end{array}$} \\
     & & $N=4$ & $\scriptstyle (\num{1.003(25)}) + (\num{0.3001(12)})|z|^4 + (\num{0.04646(18)})|z|^8 + \cdots$ &  \\
     & & \multirow{2}{*}{$N=6$} & \multirow{2}{*}{\renewcommand{\arraystretch}{1}$\begin{array}{l}\scriptstyle (\num{0.97(7)}) + (\num{0.294(016)})|z|^4 + (\num{0.0461(10)})|z|^8 \\ \scriptstyle + (\num{0.00521(4)})|z|^{12} + \cdots\end{array}$} \\
     & & & \\
     \cline{2-5}
     & \multirow{4}{*}{$4$} & $N=2$ & $1+\frac{2 |z|^{4}}{7}+\frac{3 |z|^{8}}{140} + \cdots$ & \multirow{4}{*}{$\begin{array}{l}1+\frac{2 |z|^{4}}{7}+\frac{13 |z|^{8}}{315}+\frac{2 |z|^{12}}{495} + \cdots \\ \scriptstyle {}=1 + \num{0.285714}|z|^4 + \num{0.0412698}|z|^8 + \num{0.0040404}|z|^{12} + \cdots\end{array}$} \\
     & & $N=4$ & $\scriptstyle (\num{0.91(9)}) + (\num{0.278(16)})|z|^4 + (\num{0.0412(15)})|z|^8 + \cdots$ &  \\
     & & \multirow{2}{*}{$N=6$} & \multirow{2}{*}{\renewcommand{\arraystretch}{1}$\begin{array}{l}\scriptstyle (\num{1.1(8)}) + (\num{0.243(34)})|z|^4 + (\num{0.039(5)})|z|^8 \\ \scriptstyle + (\num{0.0041(4)})|z|^{12} + \cdots\end{array}$} \\
     & & & \\
     \cline{2-5}
     & $5$ & $N=2$ & $1+\frac{5 |z|^{4}}{18}+\frac{5 |z|^{8}}{252} + \cdots$ & $1+\frac{5 |z|^{4}}{18}+\frac{215 |z|^{8}}{5544} + \cdots$ \\
     \cline{2-5}
     & $6$ & $N=2$ & $1+\frac{3 |z|^{4}}{11}+\frac{5 |z|^{8}}{264} + \cdots$ & $1+\frac{3 |z|^{4}}{11}+\frac{16 |z|^{8}}{429} + \cdots$ \\
     \cline{2-5}
     & $7$ & $N=2$ & $1+\frac{7 |z|^{4}}{26}+\frac{21 |z|^{8}}{1144} + \cdots$ & $1+\frac{7 |z|^{4}}{26}+\frac{623 |z|^{8}}{17160} + \cdots$ \\
     \cline{2-5}
     & $8$ & $N=2$ & $1+\frac{4 |z|^{4}}{15}+\frac{7 |z|^{8}}{390} + \cdots$ & $1+\frac{4 |z|^{4}}{15}+\frac{118 |z|^{8}}{3315} + \cdots$ \\
    \hline
    \multirow{5}{*}{C} & \multirow{3}{*}{$1$} & $N=2$ & $1+\frac{ |z|^{4}}{6}$ & \multirow{3}{*}{$1+\frac{ |z|^{4}}{6}+\frac{ |z|^{8}}{120}+\frac{ |z|^{12}}{5040} + \cdots$} \\
     & & $N=4$ & $1+\frac{ |z|^{4}}{6}+\frac{ |z|^{8}}{120}$ & \\
     & & $N=6$ & $1+\frac{ |z|^{4}}{6}+\frac{ |z|^{8}}{120}+\frac{ |z|^{12}}{5040} + \cdots$ \\
     \cline{2-5}
    & \multirow{2}{*}{$2$} & $N=2$ & $1+\frac{ |z|^{4}}{5}+\frac{ |z|^{8}}{120}$ & \multirow{2}{*}{$1+\frac{ |z|^{4}}{5}+\frac{2 |z|^{8}}{105} + \cdots$} \\
    & & $N=4$ & $1+\frac{ |z|^{4}}{5}+\frac{2 |z|^{8}}{105} + \cdots$ & \\
    \hline
    \multirow{10}{*}{AI} & $1$ & $N=3$ & $1+\left(x^2+y^2\right)+\left(\frac{x^4}{2}+x^2 y^2+\frac{y^4}{2}\right)$ & $1+\left(x^2+y^2\right)+\left(\frac{x^4}{2}+x^2 y^2+\frac{y^4}{2}\right)$ \\
    & & & $+\left(\frac{x^6}{6}+\frac{x^4 y^2}{2}+\frac{x^2 y^4}{2}+\frac{y^6}{6}\right)$ & $+\left(\frac{x^6}{6}+\frac{x^4 y^2}{2}+\frac{x^2 y^4}{2}+\frac{y^6}{6}\right) + \cdots$ \\ 
    \cline{2-5}
    & $2$ & $N=3$ & $1+\left(2 x^2+\frac{2 y^2}{3}\right)+\left(2 x^4+\frac{4 x^2 y^2}{3}+\frac{2 y^4}{3}\right)$ & $1+\left(2 x^2+\frac{2 y^2}{3}\right)+\left(2 x^4+\frac{4 x^2 y^2}{3}+\frac{2 y^4}{3}\right)$ \\
    & & & $+\left(\frac{4 x^6}{3}+\frac{4 x^4 y^2}{3}+\frac{4 x^2
y^4}{3}+\frac{4 y^6}{15}\right) + \cdots$ & $+\left(\frac{4 x^6}{3}+\frac{4 x^4 y^2}{3}+\frac{4 x^2
y^4}{3}+\frac{4 y^6}{15}\right) + \cdots$\\
\cline{2-5}
    & $3$ & $N=2$ & $1+\left(3 x^2+\frac{3 y^2}{5}\right)+\left(\frac{9 x^4}{2}+\frac{9 x^2 y^2}{5}+\frac{y^4}{2}\right)$ & $1+\left(3 x^2+\frac{3 y^2}{5}\right)+\left(\frac{9 x^4}{2}+\frac{9 x^2 y^2}{5}+\frac{y^4}{2}\right)$ \\
    & & & $+\left(\frac{13 x^6}{6}+\frac{29 x^4
y^2}{10}+\frac{9 x^2 y^4}{10}+\frac{y^6}{6}\right) + \cdots$ & $+\left(\frac{9 x^6}{2}+\frac{27 x^4
y^2}{10}+\frac{3 x^2 y^4}{2}+\frac{53 y^6}{210}\right) + \cdots$ \\
\cline{2-5}
    & $4$ & $N=2$ & $1+\left(4 x^2+\frac{4 y^2}{7}\right)+\left(8 x^4+\frac{16 x^2 y^2}{7}+\frac{16 y^4}{35}\right)$ & $1+\left(4 x^2+\frac{4 y^2}{7}\right)+\left(8 x^4+\frac{16 x^2 y^2}{7}+\frac{16 y^4}{35}\right)$ \\
    & & & $+\left(\frac{17 x^6}{3}+\frac{37 x^4 y^2}{7}+\frac{39
x^2 y^4}{35}+\frac{13 y^6}{105}\right) + \cdots$ & $+\left(\frac{32 x^6}{3}+\frac{32 x^4 y^2}{7}+\frac{64
x^2 y^4}{35}+\frac{64 y^6}{315}\right) +  \cdots$ \\
\cline{2-5}
    & $5$ & $N=2$ & $1+\left(5 x^2+\frac{5 y^2}{9}\right)+\left(\frac{25 x^4}{2}+\frac{25 x^2 y^2}{9}+\frac{55 y^4}{126}\right)$ & $1+\left(5 x^2+\frac{5 y^2}{9}\right)+\left(\frac{25 x^4}{2}+\frac{25 x^2 y^2}{9}+\frac{55 y^4}{126}\right)$ \\
    & & & $+\left(\frac{35 x^6}{3}+\frac{25 x^4
y^2}{3}+\frac{85 x^2 y^4}{63}+\frac{y^6}{9}\right) + \cdots$ & $+\left(\frac{125 x^6}{6}+\frac{125 x^4
y^2}{18}+\frac{275 x^2 y^4}{126}+\frac{37 y^6}{198}\right) + \cdots$ \\
    \hline \hline
\end{tabular}
    \label{tab: comparison saddle point vs small N}
\end{table*}

%%%%%%%%%%
\section{Details on non-Hermitian random matrices}
    \label{asec: NH RMT}

In this Appendix,
in addition to class D (see Sec.~\ref{sec: RMT} for details), we derive the nonlinear sigma models for other symmetry classes and formulate them in terms of Selberg-type integrals.
We also collect the characteristic polynomials for several symmetry classes in Table~\ref{tab: nh rmt known results}.
We verify that these results are exact in the large-$N$ limit $N\to\infty$.
Additionally, 
through analytic continuation to the negative replica index $n < 0$, we also demonstrate the duality between different symmetry classes.

%%%%%%%%%%
\subsection{Nonlinear sigma models for other classes}

In Eqs.~\eqref{eq: class d saddle pt odd} and \eqref{eq: class d saddle pt even}, we show that the characteristic polynomial $Z_n$ for class D is written as a Selberg-type integral $I^{(\alpha,\beta,\gamma)}$.
Here, we derive $Z_n$ for class C and illustrate that it can be written in terms of such an integral as well.
Non-Hermitian random matrices in class C are defined to respect particle-hole symmetry in Eq.~(\ref{eq: PHS}), where a unitary matrix $\mathcal{C}$ satisfies $\mathcal{C}\mathcal{C}^{*} = -1$.
Without loss of generality, we choose $\mathcal{C}$ to be $\sigma_y$, for which the matrices satisfy $\sigma_y H^T \sigma_y = -H$.
To obtain the nonlinear sigma model for $Z_n$ in Eq.~(\ref{eq: replica partition function}), we first apply the Hubbard-Stratonovich transformation, leading to
\begin{align}
    Z_n \left( E \right) &= \int \mathcal{D}Q e^{-(1/2)\,\tr{Q^\dagger Q}} \left[ \det\begin{pmatrix}
        \left| E \right| \sigma_z & Q \\
        -Q^\dagger & \left| E \right| \sigma_z
    \end{pmatrix} \right]^{N/2}, \label{eq:class_c_color_flavor}
\end{align}
with
\begin{align}
    Q = \begin{pmatrix}
        X & Y \\ -Y^* & X^*
    \end{pmatrix}, \quad X,Y\in \mathbb{C}^{n\times n}.
\end{align}
The saddle point is given by $Q^\dagger Q = N$.
Next, we take the limit $N\to \infty$, so that the integral is restricted on the saddle-point manifold $\mathrm{Sp} \left( n \right)$, yielding
\begin{equation}
    \label{eq: class C saddle point}
    Z_n \simeq \int_{Q^\dagger Q = N} \mathcal{D}Q \left[ \det\begin{pmatrix}
        \left| E \right| \sigma_z & Q \\
        -Q^\dagger & \left| E \right| \sigma_z
    \end{pmatrix} \right]^{N/2}.
\end{equation}
The integrand is invariant under $Q\mapsto UQV$ for $U,V\in \mathrm{U} \left( n \right)$, where $\mathrm{U} \left( n \right)$ is a subgroup of $\mathrm{Sp} \left( n \right)$.
This enables us to perform a decomposition that block-diagonalizes $Q$ and reduces the matrix integral to a Selberg-type integral.
We can also use the same method as in class D by treating Eq.~\eqref{eq: class C saddle point} as the characteristic polynomial of a circular ensemble and converting the integral to a Selberg-type integral, supported by numerical evidence.
Specifically, the one-point characteristic polynomial $Z_n(E, E^*)$ in the large-$N$ limit 
is given as Eq.~\eqref{class c result}.

Following the same procedure, the characteristic polynomials $Z_n$ for classes AI and AII can also be written in terms of Selberg-type integrals~\cite{Nishigaki-02}.
For all these classes, after taking the Hubbard-Stratonovich transformation and the large-$N$ limit, $Z_n$ has a similar form to Eq.~\eqref{eq:class d saddle point}.
In Table~\ref{tab: nh rmt known results}, we list the manifold of integration and the symmetry group $G$ such that the integrand is invariant under $Q\mapsto UQV$ for $U,V\in G$;
see the columns ``Manifold'' and ``Symmetry'', respectively.
Although the decomposition similar to the singular-value decomposition is not necessarily available for all the cases, symmetry $Q\mapsto UQV$ still reduces enough degrees of freedom so that the matrix integral for $Z_n$ becomes $I^{(\alpha,\beta,\gamma)}$, where $\alpha$ and $\beta$ are obtained numerically.
We list obtained $Z_n$'s in the column ``Characteristic polynomial'' of Table~\ref{tab: nh rmt known results}.

%%%%%%%%%%
\subsection{Selberg-type integrals}

If we are interested in level-level correlations, we can also apply the replica method to the two-point characteristic polynomial,
\begin{align}
    &Z_n \left( E_1, E_2 \right) \coloneqq \langle \det \left( E_1 - H \right)^n \det \left( E_1^* - H^\dagger \right)^n \nonumber \\
    &\qquad\qquad\qquad \times \det \left( E_2 - H \right)^n \det \left( E_2^* - H^\dagger \right)^n \rangle.
\end{align}
After taking the Hubbard-Stratonovich transformation and the large-$N$ limit, $Z_n \left( E_1, E_2 \right)$ again has a similar form to Eq.~\eqref{eq:class_c_color_flavor}.
In classes A, AI$^\dagger$, and AII$^\dagger$,  $Z_n \left( E_1, E_2 \right)$ can be written in terms of Selberg-type integrals since a decomposition of $Q$, similar to the singular-value decomposition, is feasible for the integrand~\cite{Nishigaki-02, Akemann-24, Anish}.
We list these expressions for $Z_n \left( E_1, E_2 \right)$ in the rows ``2pt'' of Table~\ref{tab: nh rmt known results}.

The above results are valid for $N\to \infty$.
In classes A, $\mathrm{AI}^\dagger$, and $\mathrm{AII}^\dagger$, the one-point characteristic polynomial $Z_n \left( E \right)$ can be written in terms of Selberg-type integrals also for finite $N$.
Notably, the Hubbard-Stratonovich transformation is exact for arbitrary $N$.
Without taking the large-$N$ limit $N\to\infty$, the integrand can be decomposed in classes A, $\mathrm{AI}^\dagger$, and $\mathrm{AII}^\dagger$.
The manifolds of integration are respectively $\mathbb{C}^{n\times n}$, $\mathbb{H}^{n\times n}$, and $\mathbb{R}^{2n\times 2n}$, while the symmetry groups $G$ are $\mathrm{U} \left( n \right)$, $\mathrm{Sp} \left( n \right)$, and $\mathrm{O} \left( 2n \right)$.
After these decompositions, $Z_n \left( E \right)$ is found to be given in terms of noncompact Selberg-type integrals, defined by
\begin{align}
    &\mathcal{J}^{(\alpha,\beta,\gamma)}_k \left( t \right) \coloneqq \int_{[0,\infty]^k} \exp\left(-t\sum_{\ell=1}^k \lambda_\ell\right) \nonumber \\
    &\qquad\quad \times \left( |\Delta(\lambda)|^{2\gamma} \prod_{\ell=1}^k \lambda_\ell^{\alpha-1} (1+\lambda_\ell)^{\beta-1} d \lambda_\ell\right).
\end{align}
We also list these expressions in the rows ``$\forall\,N$" of Table~\ref{tab: nh rmt known results}.

Finally, we also have bosonic characteristic polynomials, defined by 
\begin{equation}
    Z^{\text{b}}_n \left( E \right) \coloneqq \Braket{\left[ \det \left( E-H \right) \left( E^{*} - H^{\dag} \right) \right]^{-n}}
\end{equation}
with $n > 0$.
In class AI, $Z^{\text{b}}_n \left( E \right)$ was obtained in Ref.~\cite{Nishigaki-02}.
The inverse of the determinant is given by an integral of complex numbers, instead of Grassmann numbers.
Consequently, the saddle-point manifold becomes noncompact.
After the symmetry-based decomposition, $Z^{\text{b}}_n \left( E \right)$ is written in terms of noncompact Selberg-type integrals, summarized in the row ``bosonic'' in Table~\ref{tab: nh rmt known results}.

%%%%%%
\subsection{Fermionic saddle-point integral}

In Table~\ref{tab: comparison saddle point vs small N}, we compare one-point characteristic polynomials for classes D, C, and AI computed analytically or numerically for small $N$ with those obtained through the saddle-point analysis, which is supposed to be exact for $N \to \infty$.
Even for small $N$, the terms of lowest degree already agree.
As $N$ increases, more terms align with the saddle-point results.
Therefore, the results listed in Table~\ref{tab: nh rmt known results} are exact for $n > 0$.
However, because of the inherent ambiguity in analytic continuation, a straightforward approach to taking the replica limit $n \to 0$ fails to yield the correct density of eigenvalues.

%%%%%%
\subsection{Duality between different symmetry classes}

In Eq.~\eqref{eq: class d and c duality}, we discuss the duality of non-Hermitian random matrices between classes D and C.
This arises from the duality between $\mathrm{Sp} \left( n \right)$ and $\mathrm{O} \left( -2n \right)$.
We can also find similar duality in other symmetry classes.
In classes AI and AII, $Z_n \left( E \right)$'s are related by $n \mapsto -n$ and $\left( \mathrm{Im}\,E \right)^2 \mapsto -\left( \mathrm{Im}\,E \right)^2$.
Additionally, in classes $\mathrm{AI}^\dagger$ and $\mathrm{AII}^\dagger$, the two-point characteristic polynomials $Z_n \left( E_1, E_2 \right)$ are related by $n \mapsto -n$, i.e., $Z_n^{\mathrm{AI}^\dagger} \left( E_1, E_2 \right) = Z_{-n}^{\mathrm{AII}^\dagger} \left( E_1, E_2 \right)$.
In class A, $Z_n \left( E_1, E_2 \right)$ is invariant under $n \to -n$, i.e., $Z_n^{\mathrm{A}} \left( E_1, E_2 \right) = Z_{-n}^{\mathrm{A}} \left( E_1, E_2 \right)$.

The duality of the one-point fermionic characteristic polynomials $Z_n \left( E \right)$ for classes $\mathrm{AI}^\dagger$, $\mathrm{AII}^\dagger$, and A is formulated in terms of $\gausshypergeometricjack{1/\gamma}$.
Similar to $I^{(\alpha,\beta,\gamma)}_k$ given in terms of $\confluenthypergeometricjack{1/\gamma}$, $\mathcal{J}^{(\alpha,\beta,\gamma)}_k$ is given in terms of $\Psi^{(1/\gamma)}$ by
\begin{align}
    \mathcal{J}^{(\alpha,\beta,\gamma)}_k \left( t \right)
    &=\Psi^{(1/\gamma)}\left(\begin{array}{c}
        \alpha + \gamma \left( k-1 \right) \\ \alpha+\beta+2\gamma \left( k-1 \right)
    \end{array};t^{\oplus k}\right). 
    \label{eq: noncompact hypergeometric integral}
\end{align}
Then, $Z_n \left( E \right)$'s for classes AI$^\dagger$ and AII$^\dagger$ are related to each other (up to a factor of $|E|^{n N}$) by $b \mapsto -b/2$, $N \mapsto -N/2$, $n\mapsto -n$, and $|E|^2 \mapsto -|E|^2$.
In class A, $Z_n \left( E \right)$ is invariant under $b \mapsto -b$, $N \mapsto -N$, $n \mapsto -n$, and $|E|^2 \mapsto -|E|^2$.

For $n\notin\mathbb{Z}$, these duality relations become invalid, or the characteristic polynomial $Z_n$ exhibits a singularity at $n=0$.
Otherwise, enforcing duality and smoothness of $Z_n$ at $n=0$ leads to incorrect results, such as the negative density of states.

%\clearpage
\onecolumngrid
%%%%%%%%%%%%
\section{Hypergeometric functions}
    \label{asec: hypergeometric}

We begin with reviewing notations related to symmetric polynomials.
$\mathcal{P} \left( n \right)$ is defined as the set of partitions of $n\in \mathbb{Z}$.
We have the following definitions associated with $\mu\in\mathcal{P} \left( n \right)$:
$C^{(1/\gamma)}_\mu \left( t_1,\cdots,t_k \right)$ is the Jack polynomial for $\mu\in\mathcal{P} \left( n \right)$, 
which is a homogeneous polynomial of $(t_1,\cdots,t_k)$ of degree $n$;
$(a)_\mu$ is the generalized Pochhammer symbol, defined as
\begin{equation}
    (a)^{(1/\gamma)}_\mu \coloneqq \prod_{i=1}^n \prod_{j=1}^{\mu_i} \left( a-\gamma \left(i-1 \right)+\left( j-1 \right) \right).
\end{equation}

Next, we summarize the properties of confluent hypergeometric functions associated with the Jack polynomials and their relations to the integrals used in this work.
Like the single-variable counterparts, there are two kinds of confluent hypergeometric functions associated with the Jack polynomials, denoted by $\confluenthypergeometricjack{1/\gamma}$ and $\tricomihypergeometricjack{1/\gamma}$.
We have a few equivalent manners to define these functions, as follows~\cite{Yan1992, Kaneko1993}.

%\onecolumngrid
First, from the series expansion,  $\gausshypergeometricjack{1/\gamma}$ can be defined as
%\begin{align}
%    &\phantom{{}={}} \gausshypergeometricjack{1/\gamma}\hypergeometricparams{a,b}{c}{t_1,\cdots,t_k} \nonumber \\ 
%    &\coloneqq \sum_{j=0}^\infty \sum_{\mu\in\mathcal{P}(j)} \frac{1}{j!} \frac{(a)_\mu (b)_\mu}{(c)_\mu} C^{(1/\gamma)}_\mu \left( t_1,\cdots,t_k \right).
%\end{align}
\begin{equation}
    \gausshypergeometricjack{1/\gamma}\hypergeometricparams{a,b}{c}{t_1,\cdots,t_k}
    \coloneqq \sum_{j=0}^\infty \sum_{\mu\in\mathcal{P}(j)} \frac{1}{j!} \frac{(a)_\mu (b)_\mu}{(c)_\mu} C^{(1/\gamma)}_\mu \left( t_1,\cdots,t_k \right),
\end{equation}
and $\confluenthypergeometricjack{1/\gamma}$ can be defined as
%\begin{align}
%    &\phantom{{}={}} \confluenthypergeometricjack{1/\gamma}\hypergeometricparams{a}{c}{t_1,\cdots,t_k} \nonumber \\ 
%    &\coloneqq \sum_{j=0}^\infty \sum_{\mu\in\mathcal{P}(j)} \frac{1}{j!} \frac{(a)_\mu}{(c)_\mu} C^{(1/\gamma)}_\mu \left( t_1,\cdots,t_k \right).
%\end{align}
\begin{equation}
\confluenthypergeometricjack{1/\gamma}\hypergeometricparams{a}{c}{t_1,\cdots,t_k} 
    \coloneqq \sum_{j=0}^\infty \sum_{\mu\in\mathcal{P}(j)} \frac{1}{j!} \frac{(a)_\mu}{(c)_\mu} C^{(1/\gamma)}_\mu \left( t_1,\cdots,t_k \right).
\end{equation}
While the series expansion for $\tricomihypergeometricjack{1/\gamma}$ seems unknown, it is defined as the confluent limit of $\gausshypergeometricjack{1/\gamma}$, as its name suggests.
Specifically, just as in the single-variable case, $\confluenthypergeometricjack{1/\gamma}$ and $\tricomihypergeometricjack{1/\gamma}$ can also be defined as the limits of $\gausshypergeometricjack{1/\gamma}$:
%\begin{align}
%    &\phantom{{}={}}\confluenthypergeometricjack{1/\gamma}\hypergeometricparams{a}{c}{t_1,\cdots,t_k} \nonumber \\ 
%    &\coloneqq \lim_{b\to\infty} \gausshypergeometricjack{1/\gamma}\hypergeometricparams{a,b}{c}{\frac{t_1}{b},\cdots,\frac{t_k}{b}},
%\end{align}
%\begin{align}
%    &\phantom{{}={}}\tricomihypergeometricjack{1/\gamma}\hypergeometricparams{a}{c'}{t_1,\cdots,t_k} \nonumber \\ 
%    &\coloneqq (t_1\cdots t_k)^{-a} \lim_{c\to\infty} \gausshypergeometricjack{1/\gamma} \nonumber \\
%    &\quad \hypergeometricparams{a,a-c'+\gamma \left( k-1 \right)+1}{c}{1 - \frac{c}{t_1},\cdots,1 - \frac{c}{t_k}}.
%\end{align}
\begin{align}
    \confluenthypergeometricjack{1/\gamma}\hypergeometricparams{a}{c}{t_1,\cdots,t_k}  
    &\coloneqq \lim_{b\to\infty} \gausshypergeometricjack{1/\gamma}\hypergeometricparams{a,b}{c}{\frac{t_1}{b},\cdots,\frac{t_k}{b}}, \\
    \tricomihypergeometricjack{1/\gamma}\hypergeometricparams{a}{c}{t_1,\cdots,t_k} 
    &\coloneqq \lim_{b\to\infty} b^{-na} \gausshypergeometricjack{1/\gamma} \hypergeometricparams{a,b-a+c-\gamma \left( k-1 \right)-1}{b}{1 - \frac{t_1}{b},\cdots,1 - \frac{t_k}{b}}.
\end{align}

Furthermore, $\confluenthypergeometricjack{1/\gamma}$ and $\tricomihypergeometricjack{1/\gamma}$ satisfy the same differential equation~\cite{Muirhead1970A, *Muirhead1970B}:
%\begin{align}
%    & t_i \partial^2_i F + \left(c - \gamma \left( k - 1 \right) -t_i \right) \partial_i F \nonumber \\  
%    &\quad + \gamma \sum_{j\neq i} \frac{t_i}{t_i - t_j} \partial_i F - \gamma \sum_{j\neq i} \frac{t_j}{t_i - t_j} \partial_j F = a F,
%\end{align}
\begin{equation}
    t_i \frac{\partial^2 F}{\partial t_i^2} + \left(c - \gamma \left( k - 1 \right) -t_i \right) \frac{\partial F}{\partial t_i} + \gamma \sum_{j\neq i} \frac{t_i}{t_i - t_j} \frac{\partial F}{\partial t_i} - \gamma \sum_{j\neq i} \frac{t_j}{t_i - t_j} \frac{\partial F}{\partial t_j} = a F,
\end{equation}
where $F$ is either $\displaystyle \confluenthypergeometricjack{1/\gamma}\hypergeometricparams{a}{c}{t_1,\cdots,t_k}$ or $\displaystyle \tricomihypergeometricjack{1/\gamma}\hypergeometricparams{a}{c}{t_1,\cdots,t_k}$.
Then, $\confluenthypergeometricjack{1/\gamma}$ can be defined as the unique solution to this differential equation such that $F$ is symmetric in $(t_1,\cdots,t_k)$ and satisfies $F \left( 0,\cdots,0 \right) = 1$.
Although it is possible to write down the integral definition for generic $t_1,\cdots,t_k$, we here assume $t_1 = \cdots = t_k \eqqcolon t$ for simplicity.
Then, we have 
\begin{align}
    \confluenthypergeometricjack{1/\gamma}\hypergeometricparams{a}{c}{t, \cdots, t} 
    &= I^{(a - \gamma\,(k-1),\,c - a - \gamma\,(k-1),\,\gamma)}_k \left( t \right), \\
    \tricomihypergeometricjack{1/\gamma}\hypergeometricparams{a}{c}{t, \cdots, t} 
    &\propto \mathcal{J}^{(a - \gamma\,(k-1),\,c - a - \gamma\,(k-1),\,\gamma)}_k \left( t \right),
\end{align}
leading to the desired integrals in Eqs.~\eqref{eq: hypergeometric integral} and \eqref{eq: noncompact hypergeometric integral}.

\twocolumngrid

\bibliography{NH_NLSM.bib}

\end{document}